\documentclass[11pt]{article}
\usepackage{axodraw}
\usepackage{epsfig}
\usepackage{amsfonts}
\usepackage{amsmath}
\usepackage{bbm,bm}
\usepackage{graphicx}
\allowdisplaybreaks[1]

\input pix.sty

\newcommand{\barkn}{n}

\renewcommand{\eq}{eq.~}
\renewcommand{\eqs}{eqs.~}
\renewcommand{\se}{sec.~}
\renewcommand{\ses}{secs.~}
\renewcommand{\fig}{fig.~}

\newcommand{\tinymsbar}{{\overline{\mbox{\tiny\rm{MS}}}}}
\newcommand{\Lambdamsbar}{{\Lambda_\tinymsbar}}
\newcommand{\alphas}{\alpha_{\rm s}}
\newcommand{\Nf}{N_{\rm f}}
\newcommand{\Nc}{N_{\rm c}}

\newcommand{\Tc}{T_{\rm c}}

\newcommand{\mE}{m_\rmii{E}}
\newcommand{\gE}{g_\rmii{E}}
\newcommand{\gammaE}{\gamma_\rmii{E}}
\newcommand{\rmO}{{\mathcal{O}}}

\newcommand{\CF}{C_\rmii{F}}
\def\lsi{\raise0.3ex\hbox{$<$\kern-0.75em\raise-1.1ex\hbox{$\sim$}}}
\def\gsi{\raise0.3ex\hbox{$>$\kern-0.75em\raise-1.1ex\hbox{$\sim$}}}
\newcommand{\lsim}{\mathop{\lsi}}
\newcommand{\gsim}{\mathop{\gsi}}

\newcommand{\rmii}[1]{{\mbox{\tiny\rm{#1}}}}

\newcommand{\im}{\mathop{\mbox{Im}}}
\newcommand{\Tint}[1]{{\hbox{$\sum$}\!\!\!\!\!\!\!\int\,}_{\!\!\!\!\raise-0.9ex\hbox{$\scriptstyle{#1}$}}}
\newcommand{\Tinti}[1]{{{\Sigma}\!\!\!\!\raise0.3ex\hbox{$\int$}_\rmii{${#1}$}}}

\newcommand{\bi}{\begin{itemize}}
\newcommand{\ei}{\end{itemize}}
\newcommand{\hide}[1]{ }
\newcommand{\bsl}[1]{\,\slash\!\!\!\!{#1}\,}

\newcommand{\deltabar}{\raise-0.02em\hbox{$\bar{}$}\hspace*{-0.8mm}{\delta}}
\def\TAsc(#1,#2)(#3,#4,#5)%
{\SetWidth{2.0}\CArc(#1,#2)(#3,#4,#5)\SetWidth{1.0}}
\def\Lwidth{3}

\def\TAgl(#1,#2)(#3,#4,#5){\SetWidth{2.0}\PhotonArc(#1,#2)(#3,#4,#5){\Lwidth}%
{6.283 #3 mul 360 div #4 #5 sub #4 #5 sub mul sqrt mul Tdensity mul}%
\SetWidth{1.0}}
\def\TLgl(#1,#2)(#3,#4){\SetWidth{2.0}\Photon(#1,#2)(#3,#4){\Lwidth}
{#1 #3 sub #1 #3 sub mul #2 #4 sub #2 #4 sub mul add sqrt Tdensity mul}%
\SetWidth{1.0}}

\renewcommand{\picc}[1]{\;\parbox[c]{60pt}{\begin{picture}(60,30)(0,0)
\SetWidth{1.0}\SetScale{1.3} #1 \end{picture}}\; }
\def\Lwidth{1.3}

\newcommand{\picv}[1]{\;\parbox[c]{80pt}{\begin{picture}(80,70)(0,-5)
\SetWidth{1.0}\SetScale{1.0} #1 \end{picture}}\; }
\newcommand{\picw}[1]{\;\parbox[c]{160pt}{\begin{picture}(160,70)(-80,-35)
\SetWidth{1.0}\SetScale{1.0} #1 \end{picture}}\; }
\def\mass{\picv{%
 \Aqu(40,46)(33,210,330)%
 \Lqu(-10,30)(12,30)%
 \Lqu(68,30)(90,30)%
 \GlueArc(40,13)(32,30,150){3.0}{8}%
 \Text(40,55)[c]{$\scriptstyle \omega_n \neq 0$}%
}}
\def\vertex{\picw{%
 \Photon(-70,0)(-45,0){1.5}{4}%
 \Lqu(-45,0)(-15,19)%
 \Lqu(-15,19)(10,19)%
 \Laqu(-45,0)(-15,-19)%
 \Laqu(-15,-19)(10,-19)%
 \Gluon(-15,-19)(-15,19){-3.0}{5.0}%
 \Text(-8,0)[l]{$\scriptstyle \omega_n \neq 0$}%
 \Text(-73,0)[r]{$\scriptstyle B_\mu$}%
}}
\def\xertex{\picw{%
 \Lqu(-45,20)(-15,15)%
 \Lqu(-15,15)(15,20)%
 \Laqu(-45,-20)(-15,-15)%
 \Laqu(-15,-15)(15,-20)%
 \Gluon(-15,-15)(-15,15){-3.0}{4.0}%
 \Text(-8,0)[l]{$\scriptstyle \omega_n \neq 0$}%
 \Text(-45,26)[c]{$\scriptstyle p_n $}%
 \Text(-45,-26)[c]{$\scriptstyle q_n + \omega_n $}%
 \Text(15,26)[c]{$\scriptstyle p_n - \omega_n $}%
 \Text(15,-26)[c]{$\scriptstyle q_n $}%
}}
\makeatletter \@addtoreset{equation}{section} \makeatother
\renewcommand{\theequation}{\arabic{section}.\arabic{equation}}
\makeatletter
\renewcommand\section{\@startsection {section}{1}{\z@}%
                                   {-5.5ex \@plus -1ex \@minus -.2ex}
                                   {2.3ex \@plus.2ex}%
                                   {\normalfont\large\bfseries}}
\renewcommand\subsection{\@startsection{subsection}{2}{\z@}%
                                     {-3.25ex\@plus -1ex \@minus -.2ex}%
                                     {1.5ex \@plus .2ex}%
                                     {\normalfont\normalsize\bfseries}}
\renewcommand\thesection {\@arabic\c@section}
\renewcommand\thesubsection   {\thesection.\@arabic\c@subsection}
\renewcommand{\@seccntformat}[1]{%
\csname the#1\endcsname.\hspace{1.0em}}
\makeatother

\begin{document}

\flushbottom

\begin{titlepage}

\begin{flushright}
MITP/14-025 \\ 
 \vspace*{1cm}
\end{flushright}

\begin{centering}
\vfill

{\Large{\bf
 A relation between screening masses and real-time rates
 }}

\vspace{0.8cm}

B.B.~Brandt$^{\rm a}$,
A.~Francis$^{\rm b}$, 
M.~Laine$^{\rm c}$ 
and 
H.B.~Meyer$^{\rm b}$

\vspace{0.8cm}

$^\rmi{a}$%
{\em
        Institute for Theoretical Physics, 
        University of Regensburg,  
        93040 Regensburg, Germany\\}

\vspace*{0.3cm}

$^\rmi{b}$%
{\em
        PRISMA Cluster of Excellence, 
        Institute for Nuclear Physics, 
        Helmholtz Institute Mainz, 
        Johannes Gutenberg University Mainz, 55099 Mainz, Germany\\}

\vspace*{0.3cm}

$^\rmi{c}$%
{\em
        Institute for Theoretical Physics, 
        Albert Einstein Center, 
        University of Bern, \\
        Sidlerstrasse 5, 3012 Bern, Switzerland\\}

\vspace*{0.8cm}

\mbox{\bf Abstract}

\end{centering}
  
\vspace*{0.3cm}
 
\noindent
%
Thermal screening masses related to the conserved vector current are
determined for the case that the current carries a non-zero Matsubara
frequency, both in a weak-coupling approach and through lattice QCD. We point
out that such screening masses are sensitive to the same infrared physics as
light-cone real-time rates. In particular, on the perturbative side, the
inhomogeneous Schr\"odinger equation determining screening correlators is
shown to have the same general form as the equation implementing LPM
resummation for the soft-dilepton and photon production rates from a hot QCD
plasma. The static potential appearing in the equation is identical to that
whose soft part has been determined up to NLO and on the lattice in the
context of jet quenching. Numerical results based on this potential suggest
that screening masses overshoot the free results (multiples of $2\pi T$) more
strongly than at zero Matsubara frequency. Four-dimensional lattice
simulations in two-flavour QCD at temperatures of 250 and 340 MeV confirm the
non-static screening masses at the 10\% level. Overall our results lend
support to studies of jet quenching based on the same potential at $T \gsim
250$~MeV.
%

\vfill
 
\vspace*{1cm}
  
\noindent
April 2014

\vfill
  
\end{titlepage}


%
\section{Introduction}
\la{se:intro}

Even though 
an asymptotically free gauge theory at a temperature ($T$) 
much higher than the confinement scale is sometimes called
weakly coupled, its dynamics is non-trivial. Denoting the gauge coupling by 
$g = \sqrt{4\pi\alphas}$, such a theory possesses three 
parametrically different momentum scales~\cite{gpy}: 
$\pi T$, $g T$, and $g^2 T/\pi$, with by assumption 
$\pi T \gg g T \gg g^2 T/\pi$.  
The structure of any physical observable can be viewed
in various ways: 
\begin{itemize}

\item[(i)] In a strict {\em weak-coupling expansion}, observables
are computed in a power series in $g$. It is a consequence
of the momentum scales as mentioned above that odd powers 
and logarithms of $g$ 
appear~\cite{jk,tt} and that some of the coefficients 
are non-perturbative~\cite{linde}. 
It is also commonly believed that  
the series converges slowly unless $g$ is extremely small, 
a problem often associated
with the dynamics of the intermediate scale $gT$. 

\item[(ii)] In an {\em effective theory approach}~\cite{dr1,dr2}, 
only the ``hardest'' 
scale is treated perturbatively. It is ``integrated out'' in order to 
derive an effective low-energy description for the ``soft'' scales
$gT$ and $g^2 T/ \pi$. The dynamics of the low-energy modes is 
solved non-perturbatively, often with the help of 
``dimensionally reduced'' lattice simulations.  

\item[(iii)] In principle the most precise level is a fully non-perturbative
solution of a given problem, with methods of 
four-dimensional lattice QCD. 
A major practical limitation of this approach 
is that the simulations are carried out
in the imaginary-time formalism.
If real-time observables are to be considered, 
an analytic continuation is required, which 
in practice is ill-controlled 
(for a review, see ref.~\cite{hbm}). 

\end{itemize}

There are many phenomenologically 
interesting observables in thermal QCD, notably screening
masses and real-time rates such as
the photon and dilepton production rates
from the plasma, or the rate of ``jet quenching'' of energetic probes
passing through the plasma, which are dominantly determined by the 
soft scale $gT$. Given the
systematic uncertainties of the third approach, 
it is suggestive to also follow the second approach for the study of
these observables. For screening masses related to flavour-singlet (gluonic)
states, this approach leads to a good description 
of thermal QCD down to temperatures of a few hundred MeV~\cite{mu}. 
Recently it has been proposed to apply the same approach 
to jet quenching~\cite{sch}, and indeed 
first simulation results exist already~\cite{marco}. 

Nevertheless, it may be questioned with every observable
how accurate the effective theory approach really is; certainly
it breaks down at temperatures very close to the confinement scale, 
which it does not capture. The purpose of this paper is 
to elaborate on a non-trivial if indirect crosscheck: we point 
out that there is a class of Euclidean observables, namely flavour
non-singlet (mesonic) screening masses at non-zero Matsubara 
frequency, which are sensitive to the same infrared physics 
as is relevant for jet quenching or photon and dilepton production. 
By measuring these observables
on a 4-dimensional lattice and comparing with results based on 
the effective theory approach, we can lend credibility to 
the latter.

The screening masses related to mesonic operators are at leading
order multiples of $2\pi T$, because of the boundary conditions
imposed on quarks across
the time direction. Corrections originate from a ``potential''
$V(r)\sim (g^2 T/\pi)\,\phi(gT r, g^2 T r/\pi)$. The potential balances against
a kinetic energy $\sim (1/\pi T) \partial_r^2$, so that the typical 
momentum scale probed is $1/r \sim \sqrt{g^2 T^2}\sim gT$. Therefore it would
be helpful to determine the function $\phi$ without recourse 
to any expansion, and this is what can be achieved with the second approach. 

The plan of this paper is the following. 
After defining the correlators in \se\ref{se:basic}, 
we compute them in non-interacting QCD in \se\ref{se:full_lo}. 
In \se\ref{se:eff_lo} we show that the QCD results can be reproduced
through an effective theory. The parameters of the effective theory
are determined through matching computations in \se\ref{se:vertex}, 
and in \se\ref{se:Seq} we recall how the solution of the problem within
the effective theory reduces to a 2-dimensional Schr\"odinger equation. 
Numerical estimates following from this
equation are displayed in \se\ref{se:numerics}. 
A lattice calculation in two-flavor QCD is 
presented in \se\ref{se:lattice}, where we also compare 
with the predictions following from the effective Schr\"odinger equation. 
An outlook and conclusions are offered in \se\ref{se:concl}. 

%
\section{Basic definitions}
\la{se:basic}

Letting $\gamma_\mu$ denote Euclidean Dirac matrices, 
with $\{ \gamma_\mu , \gamma_\nu \} = 2 
\delta^{ }_{\mu\nu}$ and $\gamma_\mu^\dagger = \gamma_\mu$, 
we consider the quark-connected (or flavour non-singlet)
vector current correlator
\be
 G^{(k_n)}_{\mu\nu}(z) \equiv
 \int_0^{1/T} \! {\rm d}\tau \, e^{i k_n \tau} \int_{\vec{x}}
 \Bigl\langle 
  (\bar\psi \gamma_\mu \psi)(\tau,\vec{x},z)
  (\bar\psi \gamma_\nu \psi)(0) 
 \Bigr\rangle_\rmi{c}
 \;, \la{G_z_def}
\ee
where $k_n \equiv 2\pi n T$ is a bosonic Matsubara frequency, 
$T$ is the temperature, and 
$\vec{x} \equiv (x_1,x_2)$ denotes a 2-dimensional vector in 
a ``transverse'' plane. A corresponding Fourier transform is 
formally defined as
\be
 G^{(k_n)}_{\mu\nu}(k_3) \equiv 
 \int_{-\infty}^{\infty} \! {\rm d}z \, e^{i k_3 z} \, 
  G^{(k_n)}_{\mu\nu}(z)
 \;. \la{G_k3_def}
\ee
It is also convenient to define a ``spectral function'' as 
\be
 \rho^{(k_n)}_{\mu\nu}(\omega) \equiv
 \im  G^{(k_n)}_{\mu\nu}(k_3\to -i [\omega + i 0^+])
 \;.  \la{spectral}
\ee
For $\mu = \nu$, 
$G^{(k_n)}_{\mu\nu}(z)$ is symmetric in $z\to -z$, so that 
$G^{(k_n)}_{\mu\nu}(k_3)$ is even  
and $\rho^{(k_n)}_{\mu\nu}(\omega)$ is odd in its argument. 
Then
$G^{(k_n)}_{\mu\nu}(z)$ can be represented as a Laplace transform: 
\ba
   G^{(k_n)}_{\mu\nu}(z)
  \;  \stackrel{\mu = \nu}{=} \; 
  \int_0^\infty \! \frac{{\rm d}\omega}{\pi} \, 
  e^{-\omega |z|} \, \rho^{(k_n)}_{\mu\nu}(\omega)
  \;. \la{reconstruct} 
\ea
The low-lying spectrum of $\rho^{(k_n)}_{\mu\nu}(\omega)$ 
is discrete; the corresponding energies, leading to an 
exponential falloff of $G^{(k_n)}_{\mu\nu}(z)$, 
are called screening masses. 

Not all of the components of $G^{(k_n)}_{\mu\nu}$ are independent.
Ward identities related to current conservation, 
$
 k_n G^{(k_n)}_{00} + k_3 G^{(k_n)}_{30} = 0 
$
and 
$
 k_n G^{(k_n)}_{03} + k_3 G^{(k_n)}_{33} = 0 
$,
as well as the definition of a ``longitudinal'' correlator 
$
 G^{(k_n)}_L \equiv G^{(k_n)}_{00} + G^{(k_n)}_{33}
$
which plays a role in dilepton production, lead to  
\be
  G^{(k_n)}_L(k_3) = \frac{k_n^2 + k_3^2}{k_3^2}\, G^{(k_n)}_{00}(k_3) 
 \;. \la{G_L}
\ee
It is therefore sufficient to compute $G^{(k_n)}_{00}$, 
whose analysis turns out to be simpler than that of $G^{(k_n)}_{33}$
(cf.\ ref.~\cite{agz_m}). 
Apart from $G^{(k_n)}_{00}$, we also consider the transverse part
\be
 G^{(k_n)}_{T}(k_3) \equiv \sum_{i=1}^{2} G^{(k_n)}_{ii}(k_3)
 \;, \la{G_T_def}
\ee
which is not constrained by Ward identities. 

Given that we have chosen a particular direction ($z$) in which to 
measure the correlators, it is convenient to choose a representation
of the Dirac matrices which is commensurate with this choice. 
Starting with the standard (Euclidean) representation, this can be
achieved through a transformation $\gamma_\mu \to U \gamma_\mu U^{-1}$, 
with a matrix $U$ given in ref.~\cite{nrqcd3}. After this 
transformation, the matrices $\gamma_0\gamma_\mu$ relevant for the 
``non-relativistic'' effective description 
(cf.\ e.g.\ \eq\nr{4quark}) read
\be
 \gamma_0 = \biggl( \begin{array}{rr} 0 & 1 \\ 1 & 0 \end{array} \biggr)
 \;, \quad
 \gamma_0^2 = \biggl( \begin{array}{rr} 1 & 0 \\ 0 & 1 \end{array} \biggr)
 \;, \quad
 \gamma_0\gamma_i = \epsilon_{ij}
 \biggl( \begin{array}{rr} 0 & -\sigma_j \\ \sigma_j & 0 \end{array} \biggr)
 \;, \quad
 \gamma_0\gamma_3 = 
 \biggl( \begin{array}{rr} i & 0 \\ 0 & -i \end{array} \biggr)
 \;,
 \la{dirac}
\ee
where the blocks are $2\times 2$-matrices, $\sigma_j$
are Pauli matrices, and $\epsilon_{12} = 1$.  
Unless stated otherwise, 
latin indices take values labelling the transverse directions, 
$i,j \in \{ 1,2 \}$.

In a previous study~\cite{nrqcd3}, the screening masses 
of $G_{T}^{(k_n)}$ at $k_n = 0$, as well as similar 
results for scalar and pseudoscalar densities 
and the axial current, were determined up to 
next-to-leading order (NLO). All of the screening masses are 
equal in this approximation: $m = 2\pi T + c g^2\Nc T/(2 \pi)$, 
where $c$ is a small positive coefficient whose value depends on 
the number of dynamical fermions. Numerical 
measurements (cf.\ refs.~\cite{lat0,lat1,lat2} and references therein)
have detected discrepancies with 
respect to this prediction, particularly for the scalar 
and pseudoscalar channels where the results are clearly
below $2\pi T$. Here we extend the study
to $k_n \neq 0$, whereby the coefficient $c$ and the quality 
of the comparison both change.  

%
\section{Leading-order computation in full QCD}
\la{se:full_lo}

Before considering NLO corrections, we work out 
the leading-order (LO) predictions. It turns out that analytic 
results can be given for the case that no average over the 
transverse directions is taken in \eq\nr{G_z_def}. Let us 
denote such correlators by 
\be
 G^{(k_n)}_{\mu\nu}(\vec{r}) \equiv
 \int_0^{1/T} \! {\rm d}\tau \, e^{i k_n \tau} 
 \Bigl\langle
   (\bar\psi \gamma_\mu \psi)(\tau,\vec{r})
   (\bar\psi \gamma_\nu \psi)(0) 
 \Bigr\rangle_\rmi{c}
 \;, \quad
 \vec{r} \equiv (\vec{x},z)
 \;. \la{G_r_def}
\ee
The correlators can be computed with
the mixed coordinate space-momentum space
techniques introduced in 
ref.~\cite{az}. In coordinate space, 
spatial propagators have the form 
\be
 \int \! \frac{{\rm d}^3\vec{p}}{(2\pi)^3} 
 \frac{e^{i \vec{p}\cdot\vec{r}}}{ p_n^2 + p^2} = \frac{e^{- |p_n|r}}{4\pi r}
 \;, \quad
  r \equiv |\vec{r}| 
 \;.
\ee
Subsequently one is faced with sums of the type 
\be
 \sum_{\{ p_n \}} e^{-|p_n| r - |p_n - k_n| r} \, P_\alpha(|p_n|)
 \;, 
\ee
where $\{ p_n \}$ denotes a fermionic Matsubara frequency and $P_\alpha$
is a polynomial of degree $\alpha \in \{0,1,2\} $. The sums can 
be carried out in analytic form, cf.\ e.g.\ ref.~\cite{ms}. 
Denoting 
\be
 \bar{r} \equiv 2 \pi T r \;, \quad
 \frac{ k_n }{ 2 \pi T } \, = \, n
 \;, 
\ee
we obtain (here $i,j\in\{1,2,3\}$)
\ba
 - \frac{ r^2 G^{(k_n)}_{00}(\vec{r}) }{\Nc T^3 e^{-|k_n|r }}  & = & 
  \frac{|\barkn|}{6}
 + \frac{|\barkn|^3}{3}
 + \frac{|\barkn|^2}{\bar{r}}
 + \frac{|\barkn|}{\bar{r}^2}
 + \frac{|\barkn|}{\bar{r}\sinh\bar{r}}
 + \frac{\cosh\bar{r}}{\bar{r}\sinh^2\bar{r}}
 + \frac{1}{\bar{r}^2\sinh\bar{r}} 
 \;,
 \hspace*{9mm}
 \la{G00_LO_r} \\ 
 \frac{ r^2 G^{(k_n)}_{ij}(\vec{r}) }{\Nc T^3 e^{-|k_n|r }} & = & 
 \frac{r_i r_j}{r^2}
 \biggl( 
    \frac{|\barkn|}{6}
 + \frac{|\barkn|^3}{3}
 + \frac{|\barkn|^2}{\bar{r}}
 + \frac{|\barkn|}{\bar{r}^2}
 + \frac{|\barkn|}{\bar{r}\sinh\bar{r}}
 + \frac{\cosh\bar{r}}{\bar{r}\sinh^2\bar{r}}
 + \frac{1}{\bar{r}^2\sinh\bar{r}} 
 \biggr)
 \nn  
 & - &
 \biggl( \delta_{ij} -  \frac{r_i r_j}{r^2}
 \biggr) 
 \biggl( 
  \frac{|\barkn|^2}{\bar{r}}
 + \frac{|\barkn|}{\bar{r}^2}
 + \frac{|\barkn|}{\bar{r}\sinh\bar{r}}
 + \frac{\cosh\bar{r}}{\bar{r}\sinh^2\bar{r}}
 + \frac{1}{\bar{r}^2\sinh\bar{r}} 
 \nn & & \hspace*{2cm}
 +\, \frac{|\barkn|\cosh\bar{r}}{\sinh^2\bar{r}}
 + \frac{1}{2\sinh\bar{r}}
 + \frac{1}{\sinh^3\bar{r}} 
 \biggr)
 \;. \la{G11_LO_r}
\ea
Structures with $\sinh\bar{r}$ in the denominator 
are exponentially suppressed at $\bar{r} \gg 1$; however they are
relevant for $n = 0$ in which case the other terms disappear. 
(For $n = 0$ a similar expression for the pseudoscalar
correlator was given in ref.~\cite{ff}. NLO corrections could be 
worked out with the techniques introduced in ref.~\cite{Bulk_rdep}.)

Let us now take the transverse averages $\int_\vec{x}$. The powerlike terms
can be integrated in terms of the exponential integral
\be
 E^{ }_1(z) \equiv \int_z^\infty \! {\rm d}t \, \frac{e^{-t}}{t}
 \; \stackrel{z\gg 1}{\approx} \;
 \frac{e^{-z}}{z} \biggl(1 - \frac{1}{z} + \frac{2}{z^2} + \ldots \biggr)
 \;, \la{E1}
\ee
yielding ($ \bar{z} \equiv 2 \pi T z $)
\ba
 - \frac{G^{(k_n)}_{00}(z) }{2\pi \Nc T^3}  & = & 
 e^{- |n \bar{z}| }\, \frac{ \barkn^2   }{2 |\bar{z}|} 
  \biggl( 1 + \frac{1}{|n \bar{z}|} \biggr) 
 +  E^{ }_1 (|n \bar{z}|)\, \frac{   |\barkn|(1 - \barkn^2) }{6} 
 + \rmO\Bigl(e^{- (|\barkn| + 1)|\bar{z}| }\Bigr)
 \;, \hspace*{5mm} \la{G00_LO_z}
 \\
 \frac{G^{(k_n)}_{T}(z) }{2\pi \Nc T^3}  & = & 
   e^{- |n \bar{z}| }\, \biggl[ 
   \frac{|\barkn| (\barkn^2 - 1) (1 - |n \bar{z}|) }{12}
  - \frac{\barkn^2  }{2 |\bar{z}|} 
  \biggl( 1 + \frac{1}{|n \bar{z}|} \biggr) 
 \biggr]
 \nn & + &  
 E^{ }_1 (|n \bar{z}|)\, \biggl[ 
 \frac{| \bar{z}^2 \barkn^3| (\barkn^2 - 1) }{12}
  + \frac{ |\barkn|}{6} + \frac{|\barkn^3|}{3}   
 \biggr]
 + \rmO\Bigl(e^{- (|\barkn| + 1)|\bar{z}| }\Bigr)
 \;. \la{G11_LO_z}
\ea
The equations simplify greatly for $|\barkn| = 1$
(we also assume $z > 0$ here): 
\ba
 G^{(\pm k_1)}_{00}(z)\!\! & = & \!\! -\Nc T^2 \,\frac{ e^{-\bar{z} }}{ 2 z}
 \biggl(1 + \frac{1}{\bar{z}} \biggr) + \rmO(e^{-2\bar{z}})
 \;, \la{G00_LO_asy} \\ 
 G^{(\pm k_1)}_{T}(z) \!\! & = & \!\! - \Nc T^2 \biggl[ 
 \frac{ e^{-\bar{z} }}{ 2 z}
 \biggl(1 + \frac{1}{\bar{z} } \biggr) - \pi T E^{ }_1(\bar{z})
 \biggr]
 + \rmO(e^{-2\bar{z} })
 \approx  -  \frac{\Nc T  e^{-\bar{z} }}{ 2 \pi z^2}
 \;. \hspace*{8mm} \la{G11_LO_asy} \la{G11_LO_asy2}
\ea
In order to gain an intuitive understanding, 
\eqs\nr{G00_LO_asy}, \nr{G11_LO_asy}
can be represented by spectral functions like
in \eq\nr{spectral}. We obtain, for $\omega > 0$, 
\ba
 \rho^{(k_1)}_{00}(\omega) & = & 
 -\Nc T \, \theta(\omega - k_1) \, \frac{ \omega }{4} 
 + \rmO(\theta(\omega - k_2)) 
 \;, \la{rho00_LO} \\
 \rho^{(k_1)}_{T}(\omega) & = & 
 -\Nc T \, \theta(\omega - k_1) \, \frac{ \omega^2 - k_1^2 }{4\omega} 
 + \rmO(\theta(\omega - k_2 )) 
 \;. \la{rho11_LO}
\ea
These results are reproduced below from a 
``low-energy description'', valid for the 
regime $| \omega - k_1 | \ll k_1 $, but it is already
clear that the physics corresponds to a 2-particle threshold, 
with a discontinuous ($\rho^{(k_1)}_{00}$) or 
continuous ($\rho^{(k_1)}_{T}$) spectral function. 

We note that the asymptotic 
behaviours of \eqs\nr{G00_LO_asy}, \nr{G11_LO_asy2} contain
a power-law in addition to an exponential decay. Physically, 
this corresponds to 
an approximation in which two {\em free} heavy particles are
generated with a continuous
spectrum; the extra suppression in \eq\nr{G11_LO_asy2} compared
with \eq\nr{G00_LO_asy} is due
to the fact that the latter is a $P$-channel correlator. 
After interactions are taken into account, 
the particles are bound together, and the spectrum is discrete, 
$\rho(\omega) \sim \sum_n c_n \delta(\omega - \omega_n)$; 
therefore we expect that in the full theory there is no power correction
to the exponential decay. 

In the ``static'' sector, $k_n = 0$, the roles of the two channels 
are interchanged. The spatially averaged correlators become
\ba
 G^{(0)}_{00}(z) \!\!\! & = & \!\!\!  -\Nc T^2  \biggl[ 
 \frac{ e^{-\bar{z} }}{ z}
 \biggl(1 + \frac{1}{\bar{z} } \biggr) - 2 \pi T E^{ }_1(\bar{z})  
 \biggr] + \rmO(e^{-3 \bar{z}})
  \approx  -  \frac{\Nc T   e^{-\bar{z} }}{ \pi z^2}
 \;, \hspace*{5mm} \la{G00_static_asy2} \\
 G^{(0)}_{T}(z) \!\!\! & = & \!\!\!  - \Nc T^2 \biggl[ 
 \frac{ e^{-\bar{z} }}{ z}
 \biggl(1 + \frac{1}{\bar{z} } \biggr) + 2 \pi T E^{ }_1(\bar{z}) 
 \biggr]
 + \rmO(e^{-3\bar{z} })
  \approx  -  \frac{2 \Nc T^2  e^{-\bar{z} }}{ z}
 , \hspace*{7mm} \la{G11_static_asy2}
\ea
and the corresponding spectral functions read
\ba
 \rho^{(0)}_{00}(\omega) & = & 
 -\Nc T \, \theta(\omega - k_1) \, \frac{ \omega^2 - k_1^2 }{2 \omega} 
 + \rmO(\theta(\omega - k_3)) 
 \;, \la{rho00_static} \\
 \rho^{(0)}_{T}(\omega) & = & 
 -\Nc T \, \theta(\omega - k_1) \, \frac{ \omega^2 + k_1^2 }{2 \omega} 
 + \rmO(\theta(\omega - k_3 )) 
 \;. \la{rho11_static}
\ea

%
\section{Effective description}
\la{se:eff_lo}

We now build an effective theory which allows us to describe
the physics of the correlators considered around the threshold
$\omega \sim \mathop{\mbox{max}}(k_1,k_n)$ 
(we restrict to $k_n \ge 0$ without loss of 
generality). We start with a tree-level construction, 
and promote it to loop level in \se\ref{se:vertex}. 

The correlator of \eq\nr{G_z_def} can be re-written as 
\be
 G^{(k_n)}_{\mu\nu}(z) = 
 T \int_\vec{x} \Bigl\langle 
 V^{(k_n)}_\mu(\vec{x},z) 
 V^{(-k_n)}_\nu(0) 
 \Bigr\rangle^{ }_\rmi{c}
 \;, \la{G_VV}
\ee
where 
after substituting 
 $\bar{\psi}(\tau) = 
 T \sum_{\{p_n \} } e^{-i p_n \tau} \bar{\psi}_{p_n}$, 
 ${\psi}(\tau) = 
 T \sum_{\{p_n \} } e^{i p_n \tau} {\psi}_{p_n}$, 
\be
 V^{(k_n)}_\mu(\vec{x},z) = T \sum_{\{p_n \} }
 \bar{\psi}_{p_n} (\vec{x},z)\, \gamma_\mu \, \psi_{p_n-k_n}(\vec{x},z) 
 \;. \la{Vmu}
\ee
In order to represent these operators within an effective theory, 
it is convenient to introduce an abelian source 
field $B_\mu$ which couples to \eq\nr{Vmu}. This can be 
achieved by adding  
$
 S_B \equiv \int_0^{1/T}\!{\rm d}\tau \int_{\vec{x},z} 
 \! \bar\psi \, \gamma_\mu B_\mu \psi
$
to the original QCD action, with $B_\mu$ 
expressed in Matsubara modes as
\be 
 B_\mu(\tau,\vec{x},z) \equiv \sum_{k_n} 
 B_\mu^{(k_n)}(\vec{x},z)\, e^{i k_n\tau}
 \;. 
\ee
The full action is 
$  
 S \equiv  S_\rmii{QCD} + S_B
$, 
where $S_\rmii{QCD}$ is the part without $B_\mu$. 
The vector currents and their correlators can then be derived from 
the identity
\be
 V_\mu^{(k_n)}(\vec{x},z) = \frac{\delta S_B}
 {\delta B_\mu^{(k_n)}(\vec{x},z)}
 \;. \la{defVmu}
\ee

The idea of the effective approach is dimensional reduction, 
i.e.\ keeping only
the Matsubara zero modes of the SU(3) gauge fields in the covariant 
derivatives $D_\mu = \partial_\mu - i g A_\mu$ (cf.\ ref.~\cite{hl}).
At tree-level, this means that we replace the original action through
\be
 S^{ }_\rmii{QCD} \to
 S^{ }_0 \, \equiv \int_0^{1/T}\!{\rm d}\tau \int_{\vec{x},z} 
 \! \bar\psi \, \gamma_\mu  D_\mu^{(n=0)} \psi
 \;. 
\ee 
Making use of the representation of Dirac matrices
in \eq\nr{dirac} and denoting 
\be
 \psi = \frac{1}{\sqrt{T}} 
 \left( \begin{array}{c} \chi \\ \phi \end{array} \right)
 \;, 
\ee
we thereby get
\ba
 S^{ }_0 & = & 
 \sum_{ \{ p_n \} } \int_{\vec{x},z} 
 \Bigl[
   i \chi^{\dagger}_{p_n} (p_n - g A_0 + D_3) \chi_{p_n} 
 + 
   i \phi^{\dagger}_{p_n} (p_n - g A_0 - D_3) \phi_{p_n} 
 \nn & & \qquad + \, 
 \epsilon_{ij} \Bigl(
 \chi^{\dagger}_{p_n} \sigma_i D_j \phi_{p_n}- 
 \phi^{\dagger}_{p_n} \sigma_i D_j \chi_{p_n} \Bigr)
 \Bigr]
 \;, \la{S_01} \\[2mm]
 S^{ }_B & = & 
 \sum_{\{ p_n \}, k_n  } \int_{\vec{x},z} 
 \Bigl[
   B_0^{(k_n)} 
    \Bigl( 
     \chi^{\dagger}_{p_n} \chi^{ }_{p_n - k_n } 
 + 
     \phi^{\dagger}_{p_n} \phi^{ }_{p_n - k_n}
     \Bigr) 
 +  i B_3^{(k_n)} 
    \Bigl( 
     \chi^{\dagger}_{p_n} \chi^{ }_{p_n - k_n } 
 - 
    \phi^{\dagger}_{p_n} \phi^{ }_{p_n - k_n}
     \Bigr) 
 \nn & & \qquad + \, B_i^{(k_n)}
 \epsilon_{ij} \Bigl(
 \phi^{\dagger}_{p_n} \sigma_j \chi^{ }_{p_n - k_n }- 
 \chi^{\dagger}_{p_n} \sigma_j \phi^{ }_{p_n - k_n } \Bigr)
 \Bigr]
 \;. \la{S_B}
\ea
{}From $S_0$ it is observed that free propagators, 
\be
 \langle \chi^{ }_{p_n}(z_1) \chi^\dagger_{ p_n } (z_2) \rangle \simeq
 \int_{p_3} e^{i p_3 (z_1 - z_2)} \frac{-i}{p_n + i p_3}
 \;, \quad
 \langle \phi^{ }_{p_n}(z_1) \phi^\dagger_{ p_n } (z_2) \rangle \simeq
 \int_{p_3} e^{i p_3 (z_1 - z_2)} \frac{-i}{p_n - i p_3}
 \;, \la{props}
\ee
are proportional to $\theta(z_1 - z_2)$
for $\chi^{ }_{p_n > 0}$ and $\phi^{ }_{p_n < 0}$; 
and to $\theta(z_2 - z_1)$
for $\chi^{ }_{p_n < 0}$ and $\phi^{ }_{p_n > 0}$.
For any $p_n$ one of the fields is thus ``non-propagating''
or ``short-range'' 
and can be integrated out. Given that fermionic fields appear
quadratically, the integration out can equivalently be achieved 
by solving equations of motion. This yields
the simplified representation 
\ba
 S^{ }_0 & = & 
 \sum_{ \{ p_n \} } \int_{\vec{x},z} 
 \biggl[
   i \chi^{\dagger}_{p_n} \biggl(p_n - g A_0 + D_3
   - \frac{D_iD_i + i \sigma_3 \epsilon_{ij} D_i D_j }{2p_n}
  \biggr) \chi_{p_n} 
 \nn & & \qquad + \, 
   i \phi^{\dagger}_{p_n} \biggl(p_n - g A_0 - D_3 
   - \frac{D_iD_i + i \sigma_3 \epsilon_{ij} D_i D_j }{2p_n}
  \biggr) \phi_{p_n} 
 + \rmO\biggl( \frac{1}{p_n^2} \biggr)
  \biggr]
 \;. \la{S_02}
\ea

Given that $\chi^{ }_{p_n < 0 }, \phi^{ }_{p_n > 0}$ 
are non-propagating
(we consider $z_1 > z_2$), forward-propagating mesons
are of the types $\phi^\dagger_{p_n} \chi_{p_n'}$
and $\phi^\dagger_{p_n} \phi^{ }_{-p_n'}$ with $p_n, p_n' > 0$. 
It is seen from \eq\nr{S_B} that
$B_0^{(k_n)}$ and $B_3^{(k_n)}$ couple to 
operators of this type for $0 < p_n < k_n$.  
The transverse source $B_i^{(k_n)}$ couples to 
$ 
 \epsilon_{ij} \bigl(
 \phi^{\dagger}_{p_n} \sigma_j \chi^{ }_{p_n - k_n }- 
 \chi^{\dagger}_{p_n} \sigma_j \phi^{ }_{p_n - k_n } \bigr)
$
which is non-propagating for $0 < p_n < k_n$.\footnote{%
 The mode propagates for $p_n > k_n$, but then the coefficient of the 
 exponential decay is $p_n + p_n' = 2 p_n - k_n > k_n$, i.e.\ the 
 contribution is exponentially suppressed at large distances. 
 } 
However, by making use of equations of motion for the 
non-propagating modes $\chi^{ }_{p_n - k_n }$ and $\chi^{\dagger}_{p_n}$, 
there is still a $1/p_n$ or $1/(k_n-p_n)$-suppressed projection to
a forward-propagating mode:
\ba
 && \hspace*{-0.5cm}
 V_i^{(k_n;\,p_n)} = \epsilon_{ij} \bigl(
 \phi^{\dagger}_{p_n} \sigma_j \chi^{ }_{p_n - k_n}- 
 \chi^{\dagger}_{p_n} \sigma_j \phi^{ }_{p_n - k_n} \bigr)
 \la{V_T} \\[2mm]  
 & = &  
 \phi^{\dagger}_{p_n} \biggl\{ 
 \biggl(
   \frac{1}{p_n} - \frac{1}{k_n - p_n } 
 \biggr)
 \frac{  
   \overleftrightarrow{\!{D}}_{\!\! i}\,    
  }{4 i }
 - 
 \biggl(
   \frac{1}{p_n} + \frac{1}{k_n - p_n } 
 \biggr)
\frac{   
   \sigma_3 \epsilon_{ij}
   \overleftrightarrow{\!{D}}_{\!\! j}
  }{ 4 }
 \biggr\}\, \phi^{ }_{p_n-k _n}
 + \rmO\biggl( \frac{1}{p_n}, \frac{1}{k_n - p_n } \biggr)^2
 \!, \nonumber
\ea
where 
$
    \overleftrightarrow{\!{D}}_{\!\! j}\, \equiv
    \overrightarrow{\!{\partial}}_{\!\! j}\, - 
    \overleftarrow{\!{\partial}}_{\!\! j}\, 
    - 2 i g A_j 
$,
and total derivatives were omitted. 
Therefore, the correlator $G_T^{(k_n)}$ is also non-zero; it is 
simply power-suppressed with respect to $G_{00}^{(k_n)}$.

Whereas 
the operators are of the type 
$\phi^\dagger_{p_n} \phi^{ }_{-p_n'}$ in the non-static sector, 
they are of the type $\phi^\dagger_{p_n} \chi_{p_n}$
in the static sector (i.e.\ for $k_n = 0$). For 
$V_i^{(0)}$ this is immediately visible from \eq\nr{S_B}, whereas
for $V_0^{(0)}$ the elimination of non-propagating modes 
(separately for $p_n > 0$ and $p_n < 0$) yields
\ba
 V_0^{(0;\,p_n)} = 
 \chi^{\dagger}_{p_n} \chi^{ }_{p_n} +  
 \phi^{\dagger}_{p_n} \phi^{ }_{p_n} 
 & = &  
 \frac{\epsilon^{ }_{ij}}{2 i p_n}
 \biggl\{ 
    \phi^{\dagger}_{p_n}
    \sigma^{ }_i 
    \overleftrightarrow{\!{D}}_{\!\! j}\,    
    \chi^{ }_{p_n}
    -  
    \chi^{\dagger}_{p_n}
    \sigma^{ }_i 
    \overleftrightarrow{\!{D}}_{\!\! j}\,    
    \phi^{ }_{p_n}
 \biggr\}
 + \rmO\biggl( \frac{1}{p_n^2} \biggr)
 \;.  \la{V_0} 
\ea
This is clearly a $P$-channel operator.

%
\section{Mass and vertex corrections}
\la{se:vertex}

In the discussion of the previous section, only Matsubara zero modes
of gauge fields appeared. In full QCD, there are obviously also non-zero 
Matsubara modes. The description of \eqs\nr{S_B}, \nr{S_02} should be 
viewed as a low-energy effective theory from which the non-zero 
Matsubara modes have been integrated out.
The effect of the integration out
is to modify the parameters of the low-energy description, and this is 
the topic of the present section. 

Before proceeding, let us discuss the kinematic regime
relevant for the problem. As became clear in \se\ref{se:full_lo}, 
for $k_n \neq 0$ the long-distance screening concerns a distance scale
$z \sim 1/k_n$ and is therefore determined by the kinematic regime
$K^2 = k_n^2 + k_3^2 \sim 0$. As was discussed in \se\ref{se:eff_lo}
(cf.\ e.g.\ \eq\nr{props}), the quark Matsubara modes are close to 
on-shell, with $P^2 = p_n^2 + p_3^2 \sim 0$. 
In a typical case (as discussed in more detail below)
the two ``constituents''
have the Matsubara modes $p_n = k_n/2$. Therefore, even though
we are considering a Euclidean problem, the kinematics is 
formally similar to that of collinear splitting, in which a nearly
on-shell photon with Minkowskian 
four-momentum $\mathcal{K} = (k^0,\vec{k})$ 
splits into two 
fermions with four-momenta $\mathcal{P} = \mathcal{K}/2$. 
This formal similarity suggests a relationship of the 
current problem to that of photon ($\mathcal{K}^2 = 0$) or 
soft-dilepton ($\mathcal{K}^2 \sim g^2 T^2$) 
production from a QCD plasma. 

The similarity turns out to extend into practical computations, 
notably the determination of 
effects from non-zero Matsubara modes. Indeed, 
the mass and vertex corrections induced by the non-zero modes
can be extracted from computations which are essentially equivalent to 
the derivation of the 
Hard Thermal Loop (HTL) effective action~\cite{htl1,htl2}. 
The reason
is that the assumptions needed in the computations 
are $K^2 \ll (\pi T)^2, P^2 \ll (\pi T)^2, (K-P)^2 \ll (\pi T)^2$, 
which as we have argued are true in our 
situation as well. 
The graphs to be considered are shown in \fig\ref{fig:graphs}, 
in which a 4-quark operator has been 
included as well (cf.\ ref.~\cite{hl}).

%
\begin{figure}[t]
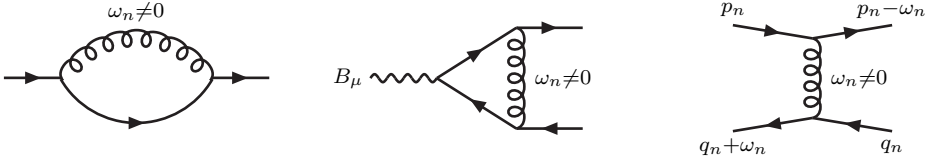


\begin{eqnarray*}
&& 
 \hspace*{5mm}
 \mass \qquad\quad \vertex \hspace*{-19mm} \xertex
\end{eqnarray*}

\caption[a]{\small 
The graphs for determining the effective mass parameter (left),  
the effective coupling of 
the vector current to the low-energy modes (middle), 
as well as 4-quark operators (right). 
} 
\la{fig:graphs}
\end{figure}
%

After computing the graphs and expanding
to leading order in $K^2/(\pi T)^2, P^2/(\pi T)^2, (K-P)^2/(\pi T)^2$, 
the results can be expressed as corrections to the actions in 
\eqs\nr{S_B}, \nr{S_02}. Using for the moment the original 
fermion fields, the free part of $S_0$ becomes\footnote{%
 In this section spatial vectors are three-dimensional 
 and latin indices run from 1 to 3.
 } 
\be
 S^{ }_0 = 
 \Tint{ \{ P \} }
 \! i\, \bar\psi(P) 
 \biggl[
   \bsl{P} +  
   \frac{m_\infty^2}{2} \int_v 
  \frac{i \gamma_0^{ } + \vec{v}\cdot\bm{\gamma}}
  {i p_n + \vec{v}\cdot\vec{p}}
 \biggr]
 \psi(P)
 \;, \la{S_0_htl}
\ee
where 
$P=(p_n,\vec{p})$, 
$\bsl{P} \equiv \gamma_\mu P_\mu$,  
$\vec{v}\cdot\bm{\gamma} \equiv v_i \gamma_i$, 
and $\int_v$ is the integral over directions of a unit vector 
($|\vec{v}| = 1$), normalized as $\int_v 1 = 1$.
The ``asymptotic mass'' parameter reads 
\be
 m_\infty^2 \equiv \frac{g^2T^2 \CF}{4}
 \;. \la{m_infty}
\ee
The coupling to the vector current is 
\be
 S^{ }_B = 
 \Tint{ \{ P,R \},K } \hspace*{-9mm}
 \bar\psi(P) 
 \biggl[
   \bsl{B}\!(K) -  
   \frac{m_\infty^2}{2} \int_v 
  \frac{(i \gamma_0^{ } + \vec{v}\cdot\bm{\gamma})
        (i B_0 + \vec{v}\cdot\vec{B})(K)}
  {(i p_n + \vec{v}\cdot\vec{p} )(i r_n + \vec{v}\cdot\vec{r} )}
 \biggr]
 \psi(R)
 \;\, \deltabar(K-P+R)
 \;. \la{S_B_htl}
\ee
It might be expected that the correction here
is suppressed by $\rmO(m_\infty^2/p_n^2) \sim \rmO(\alphas)$, 
but this is not the case, because
parts of the velocity integral give terms of 
$\rmO(m_\infty^2 / P^2) \sim \rmO(1)$. 

Let us define an ``on-shell'' spinor $u$ satisfying
\be
 \biggl[
   \bsl{P} +  
   \frac{m_\infty^2}{2} \int_v 
  \frac{i \gamma_0^{ } + \vec{v}\cdot\bm{\gamma}}
  {i p_n + \vec{v}\cdot\vec{p}}
 \biggr]
 u(P) = 0 
 \;. \la{on-shell}
\ee
Consider the dispersion relation following from \eq\nr{on-shell}. 
It is known that in Minkowskian space-time 
the dispersion relation 
of the ``particle branch'' 
reads $p^0 = p + m_\infty^2/(2 p) +  ... $, 
where $p = |\vec{p}|$~\cite{haw}. Continuing
the frequency to imaginary time, this corresponds to 
$p_n^2 + p_3^2 + \vec{p}_\perp^2 = - m_\infty^2$. 
Solving for $p_3$ with a fixed $p_n$ yields 
\be
 \pm i p_3 = p_n + \frac{m_\infty^2}{2p_n} + \frac{\vec{p}_\perp^2}{2p_n} + ...
 \;. \la{dispersion}
\ee
{}From here a ``rest mass'' can be identified and
subsequently used as a matching coefficient, 
\be
 M_n \equiv p_n + \frac{m_\infty^2}{2p_n} + \rmO(\alphas^2 T)
 \;. \la{def_Mn}
\ee
This agrees with the effective mass derived from an explicit matching
computation in ref.~\cite{nrqcd3}; a derivation through
HTL expressions like above was previously 
presented in ref.~\cite{htl_nrqcd3}. 

The computation of the vertex correction is more cumbersome; 
however, the task can be simplified  
by carrying out the matching with the special kinematics\footnote{%
 This trick can only be used if $k_n/2$ is an odd multiple of $\pi T$, 
 however we assume the result to be general. 
 } 
$R = -P = - K/2$, in an ``on-shell'' configuration.
Consider the matrix element 
\be
 \Gamma(B) \;\equiv\;
 \bar{u}(P) 
 \biggl[
   \bsl{B}\!(K) +  
   \frac{m_\infty^2}{2} \int_v 
  \frac{(i \gamma_0^{ } + \vec{v}\cdot\bm{\gamma})
        (i B_0 + \vec{v}\cdot\vec{B})(K)}
  {(i p_n + \vec{v}\cdot\vec{p} )^2}
 \biggr]
 u(-P)
 \;, \quad 
 K = 2 P
 \;. \la{GammaB}
\ee
The velocity integrals appearing here are all doable. In particular, 
it can be shown that the transverse part of the current, i.e.\ 
the part coupling to $\vec{B}^{ }_T$ with  
$\vec{p}_\perp\cdot \vec{B}^{ }_T = 0$, has a 
coefficient $\rmO(m_\infty^2/p_3^2) \sim \rmO(\alphas)$; 
this correction will be neglected in the following. 

As far as the longitudinal parts are concerned, we focus on the 
component coupling to $B_0$ like before
(cf.\ \eq\nr{G_L}). 
An explicit computation yields 
\be
 B_0 \int_v 
  \frac{i \gamma_0^{ } + \vec{v}\cdot\bm{\gamma}}
  {(i p_n + \vec{v}\cdot\vec{p} )^2}
 = -\frac{B_0}{P^2} \biggl(i \gamma_0 
 - \frac{i p_n \vec{p}\cdot\bm{\gamma}}{\vec{p}^2} \biggr)
 + \rmO\Bigl( \frac{1}{\vec{p}^2} \Bigr)
 \;. 
\ee
Inserting $P^2 = -m_\infty^2$ and $|\vec{p}_\perp| \ll |p_3| \sim |p_n|$ 
we get a correction of $\rmO(1)$, so that \eq\nr{GammaB} becomes 
\be
 \Gamma(B_0) = B_0\, \bar{u}(P) 
 \biggl[ \fr12 \biggl( \gamma_0 + \frac{p_n}{p_3} \gamma_3 \biggr)
 \biggr] u(-P) + \rmO(\alphas)
 \;. \la{Gamma_0}
\ee
Rewriting this 
with 2-component spinors like in \eq\nr{S_B}, 
the HTL-corrected vertex for the operator
to which $B_0$ couples reads
\be
 S_{B_0} \; \stackrel{p_n = k_n/2}{\to} \; \int_{\vec{x},z}
 B_0^{(k_n)}
 \biggl( \frac{p_3 + i p_n}{2 p_3} \chi^\dagger_{p_n} \chi^{ }_{-p_n}
       + \frac{p_3 - i p_n}{2 p_3} \phi^\dagger_{p_n} \phi^{ }_{-p_n} 
 \biggr)
 + \rmO(\alphas) \;. \la{S_B_final}
\ee
However, for the on-shell configuration of  
$\phi^\dagger_{p_n}$, 
$p_3 = - i p_n + \rmO(\alphas)$  (cf.\ \eq\nr{S_01}). 
Similarly, for on-shell $\chi^\dagger_{p_n}$, $p_3 =  ip_n + \rmO(\alphas)$. 
Therefore the prefactors of the operators in \eq\nr{S_B_final}
equal unity. Thus the end result is that for 
the zero component of the current, we can simply use naive
vertices as read off from \eq\nr{S_B}.\footnote{%
 It can be shown that in the static sector, $k_n = 0$, all vertex
 corrections are suppressed by $\rmO(\alphas)$, so that naive 
 vertices again suffice. 
 } 

The non-zero Matsubara modes also
induce higher-dimensional operators. In particular, as pointed out
in ref.~\cite{hl}, they generate 4-quark operators which can 
be represented as 
\be
 \delta S^{ }_0 = \frac{g^2 T}{2} \!\!\!
 \sum_{\{p_n,q_n\},\omega_n\neq 0} \! \frac{1}{\omega_n^2}
 \int_{\vec{x},z} 
 \bigl(\chi^\dagger_{p_n} \, \phi^\dagger_{p_n} \bigr)
 \, \gamma_0 \gamma_\mu \, T^a 
 \biggl( \!
  \begin{array}{c}
   \chi^{ }_{p_n-\omega_n} \\ 
   \phi^{ }_{p_n-\omega_n}
  \end{array} \! 
 \biggr) \, 
 \bigl(\chi^\dagger_{q_n} \, \phi^\dagger_{q_n} \bigr)
 \, \gamma_0 \gamma_\mu \, T^a 
 \biggl( \!
  \begin{array}{c}
   \chi^{ }_{q_n+\omega_n} \\ 
   \phi^{ }_{q_n+\omega_n}
  \end{array} \! 
 \biggr)
 \;. \la{4quark}
\ee
Here the matrices $\gamma_0\gamma_\mu$ are as given in \eq\nr{dirac}, 
and $T^a$ are Hermitean generators of SU(3), normalized as 
$\tr[T^a T^b] = \delta_{ab} / 2$.
The role of these operators is that they cause mixings; 
for instance, a state $\sim \phi^\dagger_{p_n-\omega_n} \phi^{ }_{q_n}$
can be transferred to $\sim \phi^\dagger_{p_n} \phi^{}_{q_n+\omega_n}$, 
both of which have the same screening mass $p_n - q_n - \omega_n$
at tree level, but a different ``decomposition''. This implies that
all decompositions decay with the same screening mass
when $\delta S^{ }_0$ is included.  
  
%
\section{Schr\"odinger equation}
\la{se:Seq}

In this section we recall how the computation 
of the spatial correlators within the effective theory
of \ses\ref{se:eff_lo}, \ref{se:vertex}
reduces to the solution of a two-dimensional Schr\"odinger equation.
In particular, we show that in the free limit 
\eqs\nr{rho00_LO}, \nr{rho11_LO}, \nr{rho00_static}, \nr{rho11_static} 
can be reproduced this way; 
and that, going to NLO, the equation to be solved
is closely related to that for soft-dilepton and photon
production in ref.~\cite{agz_m}. The theory is the same as
in \eq\nr{S_02}, with the modification $p_n \to M_n$ 
as discussed around \eq\nr{def_Mn}.

%
\subsection{Charge density correlator ($G^{ }_{00}$)}
\la{ss:charge}

The charge 
density can be expressed in terms of low-energy fields as 
(cf.\ \eqs\nr{defVmu}, \nr{S_B})
\be
 V_0^{(k_n)} = 
 \sum_{0 < p_n < k_n} \Bigl( 
   \chi^{\dagger}_{p_n } \chi^{ }_{p_n - k_n} + 
   \phi^{\dagger}_{p_n}  \phi^{ }_{p_n - k_n} 
  \Bigr)
 \;. 
\ee
The fields $\phi^{\dagger}_{p_n}$ 
and $\phi^{ }_{p_n-k_n} = \phi^{ }_{-|k_n - p_n|}$ are 
forward-propagating and contribute in \eq\nr{G_VV} if $z > 0$.
We now rewrite \eq\nr{G_VV} with 
an auxiliary point-splitting in the operator: 
\be
 G^{(k_n)}_{00}(z) = \lim_{\vec{y},\vec{y}'\to \vec{0}} 
 T \int_\vec{x} \Bigl\langle 
 V^{(k_n)}_0(\vec{x},z; \vec{y}) 
 V^{(-k_n)}_0(0; -\vec{y}') 
 \Bigr\rangle^{ }_\rmi{c}
 \;, \la{G00_split}
\ee
where (for $z>0$)
\be
 V_0^{(k_n)}(\vec{x},z;\vec{y}) \;\equiv\;  
 \sum_{0 < p_n < k_n}
   \phi^{\dagger}_{p_n} \! \bigl( \vec{x} + \tfr{\vec{y}}{2},z \bigr)
   \,W^{ }_{\vec{y},z}\,
   \phi^{ }_{p_n - k_n}\! \bigl( \vec{x} - \tfr{\vec{y}}{2},z \bigr)  
 \;, \la{V0_split}
\ee
and $W^{ }_{\vec{y},z}$ is a transverse Wilson line. 
Computing the correlator to leading order in the weak-coupling expansion
and taking already the limit $\vec{y}'\to\vec{0}$, 
a straightforward analysis yields
\be
  G^{(k_n)}_{00}(z) 
  = -  \sum_{0 < p_n < k_n}
  2\Nc T \lim_{\vec{y} \to \vec{0}} 
  w_\rmii{LO}(z,\vec{y})
  + \rmO(\alphas) \;, \la{G00_split_2}
\ee
where 
\be
 w_\rmii{LO}(z,\vec{y}) \equiv 
 \int_\vec{q} 
 e^{-i\vec{q}\cdot\vec{y} - ( M_\rmii{cm} + \frac{q^2}{ 2 M_\rmii{r} } ) |z| }
 \;. \la{rho_LO}
\ee
Here
\be
 M_\rmi{cm} \equiv k_n + \frac{m_\infty^2}{2M_\rmii{r}}
 \;, \quad
 M^{ }_\rmi{r} \equiv
 \biggl(
   \frac{1}{p_n} + \frac{1}{k_n - p_n}
 \biggr)^{-1}
 \;. \la{M_kn1}
\ee

Two things can be learned from \eq\nr{rho_LO}. First,  
$w_\rmii{LO}$ can equivalently be represented as
a solution of a first order differential 
equation with a particular boundary condition, 
\ba
 \Bigl( \partial_z + M_\rmi{cm} - \frac{\nabla^2}{2 M_\rmii{r}} \Bigr) 
 w_\rmii{LO}(z,\vec{y})
 & = & 0 \;, \quad z > 0 \;, 
 \la{S_lo} \\ 
 w_\rmii{LO}(0,\vec{y}) & =
 & \delta^{(2)}(\vec{y})
 \;. \la{S_init}
\ea
Second, the point-split spectral function corresponding to \eq\nr{rho_LO} can 
be determined, 
\be
 \rho_\rmii{LO}(\omega,\vec{y}) = \int_\vec{q}  
 e^{-i \vec{q}\cdot\vec{y}}\, 
 \pi \delta\Bigl( \omega - M_\rmi{cm} - \frac{q^2}{ 2 M_\rmii{r}} \Bigr)
 \;, \quad \omega > 0
 \;. \la{rho_00_split}
\ee
The original spectral function thereby becomes, combining
\eqs\nr{G00_split_2} and \nr{rho_00_split}, 
\be
 \rho^{(k_n)}_{00}(\omega) 
 \; = \; 
 - \sum_{0 < p_n < k_n} 2\Nc T \lim_{\vec{y} \to \vec{0}} 
  \rho_\rmii{LO}(\omega,\vec{y}) 
 \; = \; 
 - \sum_{0 < p_n < k_n} \Nc T
 M_\rmi{r} \, \theta \bigl( \omega - M_\rmi{cm} \bigr)
 \;. \la{rho00_nr}
\ee
Setting $n = 1$ and considering the leading order
(i.e.\ $m_\infty^2\to 0$), we have $M_\rmi{cm} = k_1$ and 
$M_\rmi{r} = k_1/4$. Then \eq\nr{rho00_nr}
agrees with \eq\nr{rho00_LO} when the latter is 
expanded to leading non-trivial order in $\omega - k_1$
(the case of general $k_n$ is discussed in appendix~A). 

Consider now NLO corrections to \eq\nr{G00_split}. 
Keeping $\vec{y},\vec{y}' \neq \vec{0}$, 
the computation can be carried out by omitting the transverse motion
suppressed by $1/(2M_\rmi{r})$, 
whereby the quark propagators are straight Wilson
lines. Sending $z \to \infty$ and suppressing $\vec{y}'$, we obtain 
\ba
 \bigl( \partial_z + M_\rmi{cm} \bigr) 
  w_\rmii{NLO}(z,\vec{y}) & \stackrel{z\to\infty}{=} &
  - V^{+}_\rmii{LO}(\vec{y})  
  w_\rmii{LO}(z,\vec{y}) 
 \;, \la{S_nlo} \\ 
 V^{+}_\rmii{LO}(\vec{y}) & \equiv & 
 \gE^2  \CF  \int_\vec{q} \Bigl( 1 - e^{i \vec{q}\cdot\vec{y}}\Bigr)
 \biggl( \frac{1}{\vec{q}^2} - \frac{1}{\vec{q}^2 + \mE^2} \biggr)
 \nn 
 & = & \frac{ \gE^2 \CF }{2\pi} 
  \biggl[ 
    \ln\Bigl( \frac{\mE y }{2} \Bigr) + \gammaE + K_0^{ }(\mE y )
  \biggr]
 \;, \la{V_nlo}
\ea
where $C_F = (\Nc^2 - 1)/(2\Nc)$; 
$\gE^2 = g^2 T$ is the gauge coupling of the dimensionally reduced theory;
$\mE^2 = (\frac{\Nc}{3} + \frac{\Nf}{6})g^2 T^2$ is 
the Debye mass parameter appearing in the static propagator of $A_0$;  
and $K_0$ is a modified Bessel function. 

We finally combine \eqs\nr{S_lo}, \nr{S_nlo}.
If we set $w_\rmii{LO} \sim \rmO(1)$, than according 
to \eq\nr{S_nlo}, $w_\rmii{NLO} \sim \rmO(\alphas)$. 
Moreover, in the kinematic regime 
$\nabla \sim gT$ of relevance to us, 
$-\nabla^2/M_\rmi{r} \! \sim \rmO(\alphas)$.
It follows that, up to a perturbative
error of $\sim\rmO(\alphas^2)$, we can write 
\be
 (\partial_z + \hat{H}^{+})w(z,\vec{y}) = 0 \;, \quad z > 0 
 \;, \la{S_full}
\ee
where $w = w_\rmii{LO} + w_\rmii{NLO} + \ldots$ and we have denoted
\be
 \hat{H}^{+} \equiv M_\rmi{cm} - \frac{\nabla^2}{2 M_\rmii{r}} + V^+ 
 \;. \la{Hplus}
\ee
The initial condition remains that same as in \eq\nr{S_init},  
up to corrections of $\rmO(\alphas)$.

The Schr\"odinger equation and the initial condition can be combined
into a single equation by taking a Fourier transform.
The system 
\be
 \partial_z {w}(z,\vec{y}) = - \mbox{sign}(z)\, \hat{H}^{+} \,
 {w}(z,\vec{y}) \;, \quad
 {w}(0,\vec{y}) = \delta^{(2)}(\vec{y}) 
\ee
can formally be solved as 
$
 {w}(z,\vec{y}) = e^{-\hat{H}^{+} |z|} {w}(0,\vec{y})
$.
Its Fourier transform  (cf.\ \eq\nr{G_k3_def}) reads
\be
 {w}(k_3,\vec{y}) = 
 \Bigl( [i k_3 + \hat{H}^{+}]^{-1} - [i k_3 - \hat{H}^{+}]^{-1}  \Bigr)
 \delta^{(2)}(\vec{y})
 \;. 
\ee
The spectral function follows from the cut. 
Defining an auxiliary function ${g}(\omega,\vec{y})$ as the solution of
a $z$-independent inhomogeneous equation
\be
 \Bigl( \hat{H}^{+} -\omega - i 0^+ \Bigr) {g}^{+}(\omega,\vec{y} ) =
 \delta^{(2)}(\vec{y}) 
 \;, \la{g_eq}
\ee
we obtain (for $\omega > 0$ and assuming a positive spectrum)
\be
 \rho^{(k_n)}_{00} (\omega) = - \sum_{0 < p_n < k_n} 2 \Nc T
 \lim_{\vec{y}\to\vec{0}} \im {g}^{+}(\omega,\vec{y})
 \;. \la{g_limit}
\ee

It may be noted that \eqs\nr{g_eq}, \nr{g_limit} bear a 
close resemblance to the corresponding equations appearing
in the LPM resummation of longitudinal modes 
for dilepton production, cf.\ 
\eqs(22), (24) of ref.~\cite{agz_m}. The overall normalizations of 
${g}^{+}$, as determined by the coefficient of the inhomogeneous
term, as well as of the parameters appearing  
do differ, but this is a matter of conventions. In addition 
some imaginary parts appear differently,\footnote{%
 In particular, in LPM resummation the potential plays the role
 of a ``width''. 
 } 
but this is 
related to the Minkowskian versus Euclidean nature of the observable
considered. The functional form of the 
potential appearing in $\hat{H}^{+}$ is identical, 
as well as the fact that we are looking for a scalar
($S$-wave) solution, as determined by the inhomogeneous term. 

%
\subsection{Transverse current correlator ($G^{ }_{T}$)}

Let us repeat the analysis for the transverse components of the current, 
cf.\ \eq\nr{G_T_def}. We again 
introduce an auxiliary point-splitting into the currents: 
\be
 G^{(k_n)}_{T}(z) = \lim_{\vec{y},\vec{y}'\to \vec{0}} 
 T  \sum_{i=1}^{2}
 \int_\vec{x} \Bigl\langle 
 V^{(k_n)}_i(\vec{x},z; \vec{y}) 
 V^{(-k_n)}_i(0; -\vec{y}') 
 \Bigr\rangle^{ }_\rmi{c}
 \;, 
\ee
where, following \eq\nr{V_T},  
\ba
 && \hspace*{-0.5cm}
 V_i^{(k_n)}(\vec{x},z;\vec{y}) 
 \equiv \sum_{0 < p_n < k_n}
 \\ 
 && 
 \phi^{\dagger}_{p_n}    \bigl(\vec{x} + \tfr{\vec{y}}{2},z\bigr) 
 \biggl\{ 
 \biggl(
   \frac{1}{p_n} - \frac{1}{k_n - p_n } 
 \biggr)
 \frac{  
   \overleftrightarrow{\!{D}}_{\!\! i}\,    
  }{4 i }
 - 
 \biggl(
   \frac{1}{p_n} + \frac{1}{k_n - p_n } 
 \biggr)
\frac{   
   \sigma_3 \epsilon_{ij}
   \overleftrightarrow{\!{D}}_{\!\! j}
  }{ 4 }
 \biggr\}\, \phi^{ }_{p_n-k _n}
 \bigl( \vec{x} - \tfr{\vec{y}}{2},z \bigr)  
 \;, \nonumber
\ea
with the notation 
$
 \overleftrightarrow{\!{D}}_{\!\! j}\, \equiv
 W^{ }_{\vec{y},z}  \overrightarrow{\!{D}}_{\!\! j}\,  - 
 \overleftarrow{\!{D}}_{\!\! j} W^{ }_{\vec{y},z} 
$. 
At leading order, 
\be
  G^{(k_n)}_{T}(z) 
 = 
 - \sum_{0 < p_n < k_n} \Nc T 
 \biggl[  \frac{1}{p_n^2} + \frac{1}{(k_n - p_n)^2} \biggr] 
 \lim_{\vec{y} \to \vec{0}} 
 \nabla \cdot \vec{v}_\rmii{LO}(z,\vec{y})
 + \rmO(\alphas) \;, \la{v_LO_1}
\ee
where we already took $\vec{y}'\to\vec{0}$ and defined 
\be
 \vec{v}_\rmii{LO}(z,\vec{y}) \equiv
 \int_\vec{q} i \vec{q}\, 
 e^{-i\vec{q}\cdot\vec{y} - ( M_\rmii{cm} + \frac{q^2}{ 2 M_\rmii{r}} ) |z| }
 \;. \la{v_LO}
\ee

Like with the charge density, the LO solution can be 
represented as a differential equation, 
\ba
 \Bigl( \partial_z + M_\rmi{cm}
 - \frac{\nabla^2}{2 M_\rmi{r} } \Bigr) 
 \vec{v}_\rmii{LO}(z,\vec{y})
 & = & 0 \;, \quad z > 0 \;, \\ 
 \vec{v}_\rmii{LO}(0,\vec{y}) & =
 & -\nabla \delta^{(2)}(\vec{y})
 \;. 
\ea
Also, a point-split 
spectral function corresponding to \eq\nr{v_LO} can be determined, 
\be
 \bm{\rho}_\rmii{LO}(\omega,\vec{y}) = \int_\vec{q} i \vec{q} \, 
 e^{-i \vec{q}\cdot\vec{y}}
 \, \pi \delta \Bigl(\omega - M_\rmi{cm} - \frac{q^2}{ 2 M_\rmii{r}} \Bigr)
 \;, \quad \omega > 0
 \;. 
\ee
The original spectral function thereby becomes
(cf.\ \eq\nr{v_LO_1})
\ba
 \rho^{(k_n)}_{T,\rmii{LO}}(\omega) & = &  
 - \sum_{0 < p_n < k_n} \Nc T  
 \biggl[  \frac{1}{p_n^2} + \frac{1}{(k_n - p_n)^2} \biggr] 
 \lim_{\vec{y} \to \vec{0}} 
 \nabla \cdot \bm{\rho}_\rmii{LO}(\omega,\vec{y}) 
 \nn 
 & = & 
 - \sum_{0 < p_n < k_n} \Nc T
 \biggl[  \frac{1}{p_n^2} + \frac{1}{(k_n - p_n)^2} \biggr] M_\rmi{r}^2
  \bigl(\omega - M_\rmi{cm} \bigr) 
  \,\theta \bigl( \omega - M_\rmi{cm} \bigr)
 \;.  \la{rhoT_nr}
\ea
The parameters appearing here are defined in \eq\nr{M_kn1}.
For $n = 1$ \eq\nr{rhoT_nr} 
agrees with \eq\nr{rho11_LO} when the latter is 
expanded to leading non-trivial order in $\omega - k_1$
(the case $n > 1$ is discussed in appendix~A). 

The inclusion of interactions proceeds like for the 
charge density, with the only difference that the ``wave function'' 
is now a vector. In particular, introducing a
$z$-independent inhomogeneous Schr\"odinger equation
\be
 \Bigl( \hat{H}^{+} - \omega - i0^+ \Bigr) \vec{f}^{+}(\omega,\vec{y} ) =
 -\nabla \delta^{(2)}(\vec{y}) 
 \;, \la{f_eq}
\ee
the cut of the solution (denoted by $\im$) yields 
the spectral function, and 
\be
 \rho^{(k_n)}_{T} (\omega) =
 - \sum_{0 < p_n < k_n} \Nc T
 \biggl[  \frac{1}{p_n^2} + \frac{1}{(k_n - p_n)^2} \biggr] 
 \lim_{\vec{y}\to\vec{0}} \im \nabla\cdot \vec{f}^{+}(\omega,\vec{y})
 \;. \la{f_limit}
\ee
Equations \nr{f_eq}, \nr{f_limit} again have the same general form as 
the ones in the LPM resummation of the photon or dilepton production rate, 
cf.\ \eqs(22), (24) of ref.~\cite{agz_m}, with 
differences originating from the chosen normalization
and parameters and 
from differences of the signatures.
The functional form of the potential is the same, as is the fact that 
we are looking for a vector-valued ($P$-wave) solution, 
as dictated by the inhomogeneous term in \eq\nr{f_eq}.

%
\subsection{Static sector}

The static case $k_n = 0$ differs qualitatively from $k_n \neq 0$. 
We go here beyond the previous discussion of ref.~\cite{nrqcd3} by including 
the charge density correlator 
and by giving the inhomogeneous Schr\"odinger 
equations determining the absolute values of the correlators.

Starting with the transverse case
(which for $n=0$ corresponds to the $S$-wave), we write
\be
 G^{(0)}_{T}(z) = \lim_{\vec{y},\vec{y}'\to \vec{0}} 
 T \sum_{i=1}^{2} \int_\vec{x} \Bigl\langle 
 V^{(0)}_i(\vec{x},z; \vec{y}) 
 V^{(0)}_i(0; -\vec{y}') 
 \Bigr\rangle^{ }_\rmi{c}
 \;, \la{GT_static_split}
\ee
where (cf.\ \eq\nr{S_B})
\be
 V_i^{(0)}(\vec{x},z;\vec{y}) \;\equiv\; 
  \sum_{\{p_n\}} \epsilon^{ }_{ij} \Bigl[ 
   \phi^{\dagger}_{p_n} \! \bigl( \vec{x} + \tfr{\vec{y}}{2},z \bigr)
   \sigma^{ }_j \,W^{ }_{\vec{y},z}\,
   \chi^{ }_{p_n} \! \bigl( \vec{x} - \tfr{\vec{y}}{2},z \bigr) - 
   (\chi\leftrightarrow\phi )
 \Bigr]
 \;. \la{Vi_static_split}
\ee
The subsequent steps go as in \se\ref{ss:charge}, 
with the difference that 
the fields appearing are $\phi^\dagger_{p_n} \chi^{ }_{p_n}$
rather than $\phi^\dagger_{p_n} \phi^{ }_{-p_n'}$, which leads to 
a different potential~\cite{nrqcd3}:
\be
 V^{-}_\rmii{LO}(\vec{y}) \,\equiv\, 
 \gE^2  \CF  \int_\vec{q} 
 \biggl( \frac{ 1 - e^{i \vec{q}\cdot\vec{y}} }{\vec{q}^2} 
 - \frac{1 + e^{i \vec{q}\cdot\vec{y}} }{\vec{q}^2 + \mE^2} \biggr)
 \, = \, \frac{ \gE^2 \CF }{2\pi} 
  \biggl[ 
    \ln\Bigl( \frac{\mE y }{2} \Bigr) + \gammaE - K_0^{ }(\mE y )
  \biggr]
 \;. \la{V_nlo_minus}
\ee
With this potential the Hamiltonian reads
\be
 \hat{H}^{-} \equiv M_\rmi{cm} - \frac{\nabla^2}{2 M_\rmi{r}} + V^{-} 
 \;, 
 \quad
 M_\rmi{cm} = 2 p_n + \frac{ m_\infty^2 }{ 2 M_\rmii{r}}
 \;, \quad
 M_\rmi{r} = \frac{p_n}{ 2} 
 \;. \la{M_kn0}
\ee
The inhomogeneous Schr\"odinger equation becomes
\be
 \Bigl( \hat{H}^{-} -\omega - i 0^+ \Bigr) {g}^{-}(\omega,\vec{y} ) =
 \delta^{(2)}(\vec{y}) 
 \;, \la{g_eq_minus}
\ee
and the spectral function is 
\be
 \rho^{(0)}_{T} (\omega) = - \sum_{p_n > 0} 8 \Nc T 
 \lim_{\vec{y}\to\vec{0}} \im {g}^{-} (\omega,\vec{y})
 \;. \la{g_limit_minus}
\ee
Here the modes $p_n > 0 $ and $p_n < 0$ have been summed 
together even though, 
when 4-quark operators are included, their degeneracy may be lifted.
In the free limit the result can be 
extracted from \eq\nr{rho00_nr}, and 
agrees with the threshold expansion of \eq\nr{rho11_static}. 

The charge density correlator
(which for $n=0$ corresponds to the $P$-wave) 
reads
\be
 G^{(0)}_{00}(z) = \lim_{\vec{y},\vec{y}'\to \vec{0}} 
 T \int_\vec{x} \Bigl\langle 
 V^{(0)}_0(\vec{x},z; \vec{y}) 
 V^{(0)}_0(0; -\vec{y}') 
 \Bigr\rangle^{ }_\rmi{c}
 \;, \la{G00_static_split}
\ee
where from \eq\nr{V_0},  
\be
 V_0^{(0)}(\vec{x},z;\vec{y}) \;\equiv\; 
  \sum_{\{p_n\}} \frac{\epsilon^{ }_{ij}}{2 i p_n} \Bigl[ 
   \phi^{\dagger}_{p_n} \! \bigl( \vec{x} + \tfr{\vec{y}}{2},z \bigr)
   \sigma^{ }_i 
   \overleftrightarrow{\!{D}}_{\!\! j}\, 
   \chi^{ }_{p_n} \! \bigl( \vec{x} - \tfr{\vec{y}}{2},z \bigr) - 
   (\chi\leftrightarrow\phi )
 \Bigr]
 \;. \la{V0_static_split}
\ee
The inhomogeneous Schr\"odinger equation reads
\be
 \Bigl( \hat{H}^{-} - \omega - i0^+ \Bigr) \vec{f}^{-}(\omega,\vec{y} ) =
 -\nabla \delta^{(2)}(\vec{y}) 
 \;, \la{f_eq_minus}
\ee
and the spectral function is 
\be
 \rho^{(0)}_{00} (\omega) = - \sum_{p_n > 0} \frac{4 \Nc T}{p_n^2} 
 \lim_{\vec{y}\to\vec{0}} \im \nabla\cdot \vec{f}^{-}(\omega,\vec{y})
 \;. \la{f_limit_minus}
\ee
Again the modes $p_n > 0 $ and $p_n < 0$ have been 
summed together. In the free limit, the result can be 
extracted from \eq\nr{rhoT_nr}, and  
agrees with the threshold expansion of \eq\nr{rho00_static}. 

%
\section{Non-perturbative potential and numerical predictions}
\la{se:numerics}

The potential given in \eq\nr{V_nlo} 
can be defined more generally from point-split
correlators such as \eq\nr{G00_split}. 
It can be defined in the ``infinite-mass'' limit
($\nabla^2/M_\rmi{r} \! \ll M_\rmi{cm}$), whereby the propagators
are just straight lines. In this situation we can set $\vec{y}' = \vec{y}$
and $\vec{x} = \vec{0}$ in \eq\nr{G00_split}. 
More specifically, let us define a loop ($z>0$)
\ba
 L(\vec{y},z) & \equiv & 
 \lim_{M_\rmii{cm} \to \infty}
 e^{M_\rmii{cm} z}
 \Bigl\langle
    V_0^{(k_n;p_n)}(\vec{0},z;\vec{y})
    \; 
    V_0^{(-k_n;-q_n)}(0;-\vec{y})
 \Bigr\rangle
 \nn 
 & = & - \tr \bigl\langle U_2^\dagger \bigl( \tfr{\vec{y}}{2},z \bigr) \,
                          W_{\vec{y},z} \,
                          U_1^{ } \bigl( - \tfr{\vec{y}}{2},z \bigr) \,
                          W^\dagger_{\vec{y},0}
  \bigr\rangle 
 \;, \la{defL}
\ea
where we inserted \eq\nr{V0_split}, 
restricted to a contribution from a single $p_n$,  
and defined 
the ``longitudinal'' Wilson lines as (${q_n} \equiv {k_n - p_n}$)
\ba
 U^{ }_1\bigl( - \tfr{\vec{y}}{2},z \bigr) & \equiv &
 \lim_{M_{q_n } \to \infty}
 e^{M_{q_n} z}
 \bigl\langle
 \phi^{ }_{- q_n}\bigl( - \tfr{\vec{y}}{2},z \bigr) 
 \phi^\dagger_{- q_n}\bigl( -\tfr{\vec{y}}{2},0 \bigr)
 \bigr\rangle^{ }_{A}
 \;, \\ 
 U^{\dagger }_2\bigl( \tfr{\vec{y}}{2},z \bigr) & \equiv &
 \lim_{M_{p_n}  \to \infty}
 e^{M_{p_n} z}
 \bigl\langle 
   \phi^{ }_{p_n}\bigl( \tfr{\vec{y}}{2},0 \bigr)
   \phi^\dagger_{p_n} \bigl( \tfr{\vec{y}}{2},z \bigr)
 \bigr\rangle^{ }_{A} 
 \;.
\ea
Here $\langle ... \rangle^{ }_A$ denotes a fermion propagator
in a fixed gauge field background. 
The Wilson lines satisfy the equations of motion
\ba
  \partial_z U^{ }_1 \bigl( - \tfr{\vec{y}}{2},z \bigr) & = &  
  \bigl(  i g A_3 - g A_0 \bigr)\, 
 U^{ }_1 \bigl( - \tfr{\vec{y}}{2},z \bigr) 
 \;, \\
 \partial_z U^{\dagger}_2 \bigl( \tfr{\vec{y}}{2},z \bigr) & = &  
 U^{\dagger }_2 \bigl( \tfr{\vec{y}}{2},z \bigr) 
 \bigl(- i g A_3 + g A_0 \bigr)\, 
 \;, 
\ea
which can be integrated in terms of path-ordered exponentials 
as usual. Note that $L(\vec{y},z)$ is $z$-independent at $\vec{y}=\vec{0}$.
The potential is subsequently extracted from 
\be
 V^+ ( \vec{y} ) \equiv 
  - \lim_{z\to\infty} L^{-1}(\vec{y},z) 
   \partial_z L(\vec{y},z)
 \;. \la{V_np}
\ee
This potential vanishes for $\vec{y} = \vec{0}$. 
The choice of the transverse Wilson lines $W,W^\dagger$
affects the overall value of $L$ but not that of the exponential
falloff, which can be viewed as an eigenvalue of a Hamiltonian. 

The potential of 
\eq\nr{V_np} agrees with the one derived and computed up to 
$\rmO(\alphas^{3/2})$
in ref.~\cite{sch}.
A coordinate space expression was given in ref.~\cite{ak}. 
The potential has 
been measured within a dimensionally reduced
effective field theory (EQCD) in ref.~\cite{marco};
we make use of the ``cold'' 
($T\approx 400$~MeV) $\beta = 16$ data set.\footnote{%
 At present no continuum extrapolation exists, but the 
 necessary ingredients have been discussed~\cite{mdo}.
 }
At short distances, a polynomial interpolation is employed;  
for estimating the potential at distances larger than 
those for which measurements exist, we  
fit the 5 largest distances ($y \gE^2 > 2.5$)
to the confining form  
$
  \sigma y + \mu + \frac{\gamma}{y}
$ which describes the asymptotics well~\cite{Cperp}.\footnote{%
 However at very large $y$, when the value of $V^+$
 exceeds $2\pi T$, we should expect string breaking
 to set in.
 } 

\begin{figure}[t]


\centerline{%
 \epsfysize=7.5cm\epsfbox{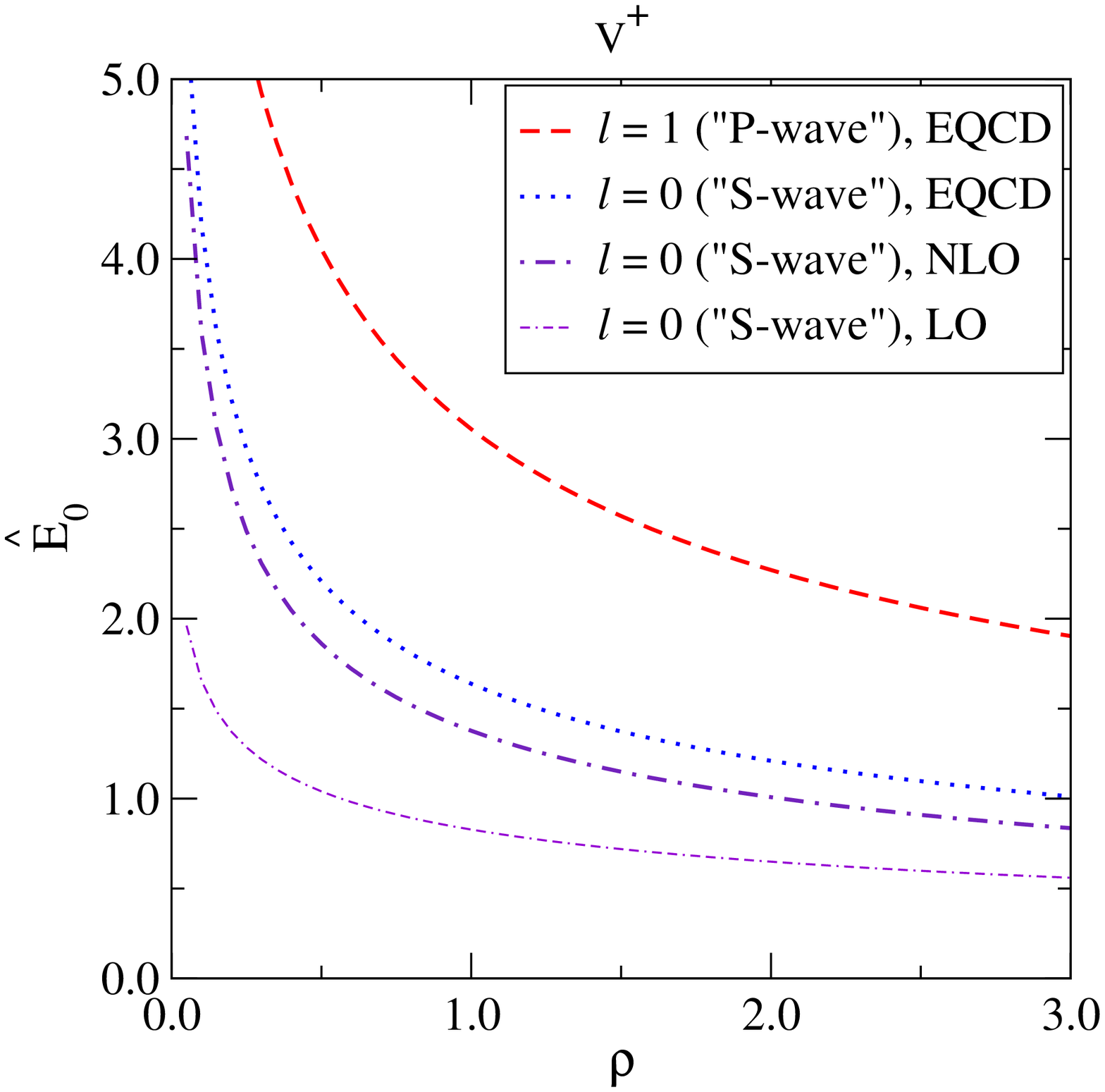}%
~~~\epsfysize=7.5cm\epsfbox{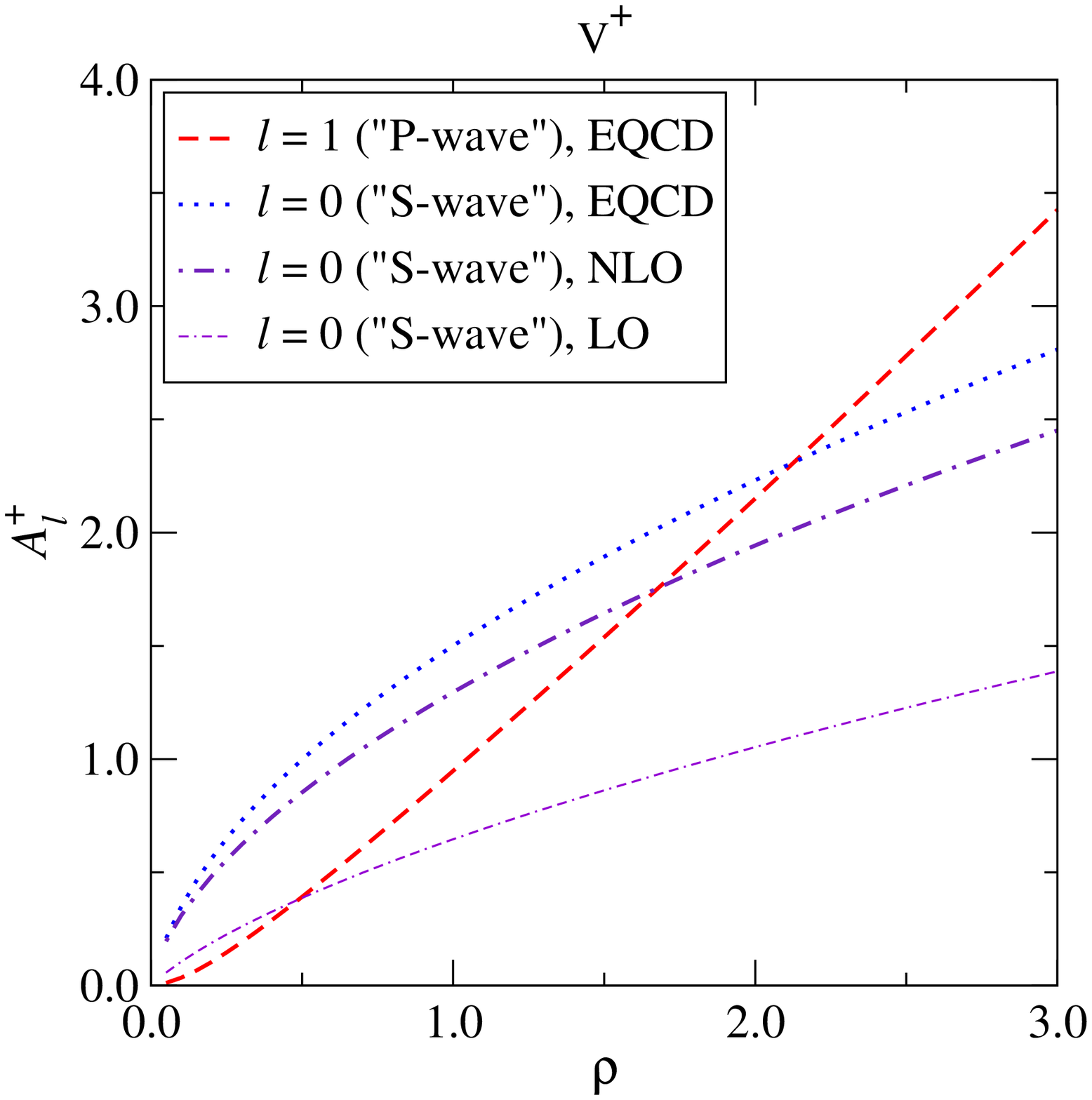}
}

\caption[a]{\small
Left: 
The lowest ``$S$-wave'' eigenvalue obtained with 
the LO (\eq\nr{V_nlo}), 
NLO~\cite{sch}, and EQCD potential $V^+$~\cite{marco}. 
For EQCD the ``$P$-wave'' result
is shown as well. For the LO case the leading-log
asymptotics read
$\hat{E}^{ }_0 \approx \fr12\! \ln \fr1{\rho}$ for $\rho \ll 1$ 
and 
$\hat{E}^{ }_0 \approx (\frac{\ln\rho}{2\rho})^{\fr12}$ for $\rho \gg 1$. 
Right:
The ``amplitudes'' corresponding to the lowest eigenmodes, 
as defined in the text (cf.\ \eqs\nr{A_0}, \nr{A_1}). 
}

\la{fig:E0hat}
\end{figure}

\begin{figure}[t]


\centerline{%
 \epsfysize=7.5cm\epsfbox{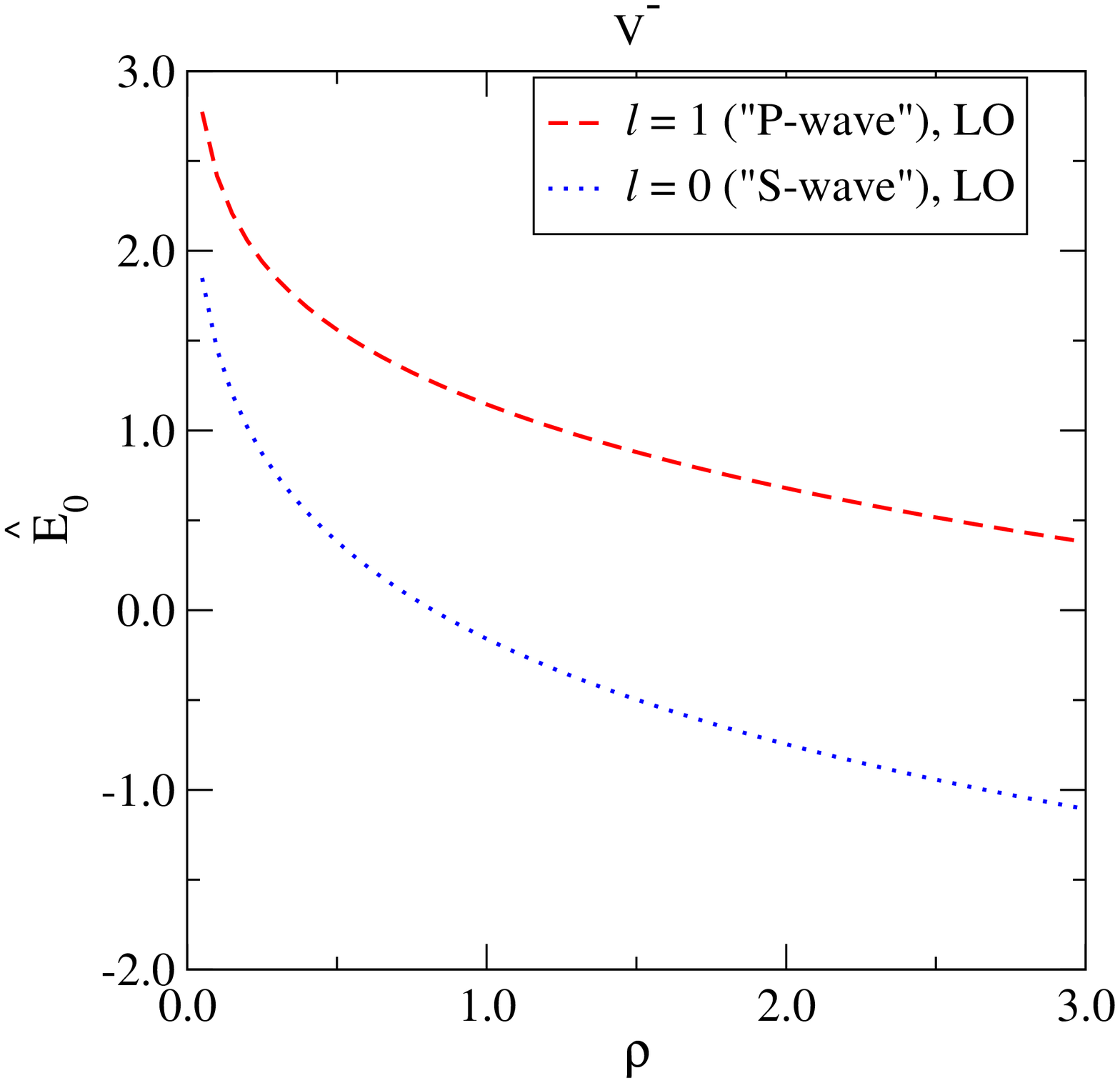}%
~~~\epsfysize=7.5cm\epsfbox{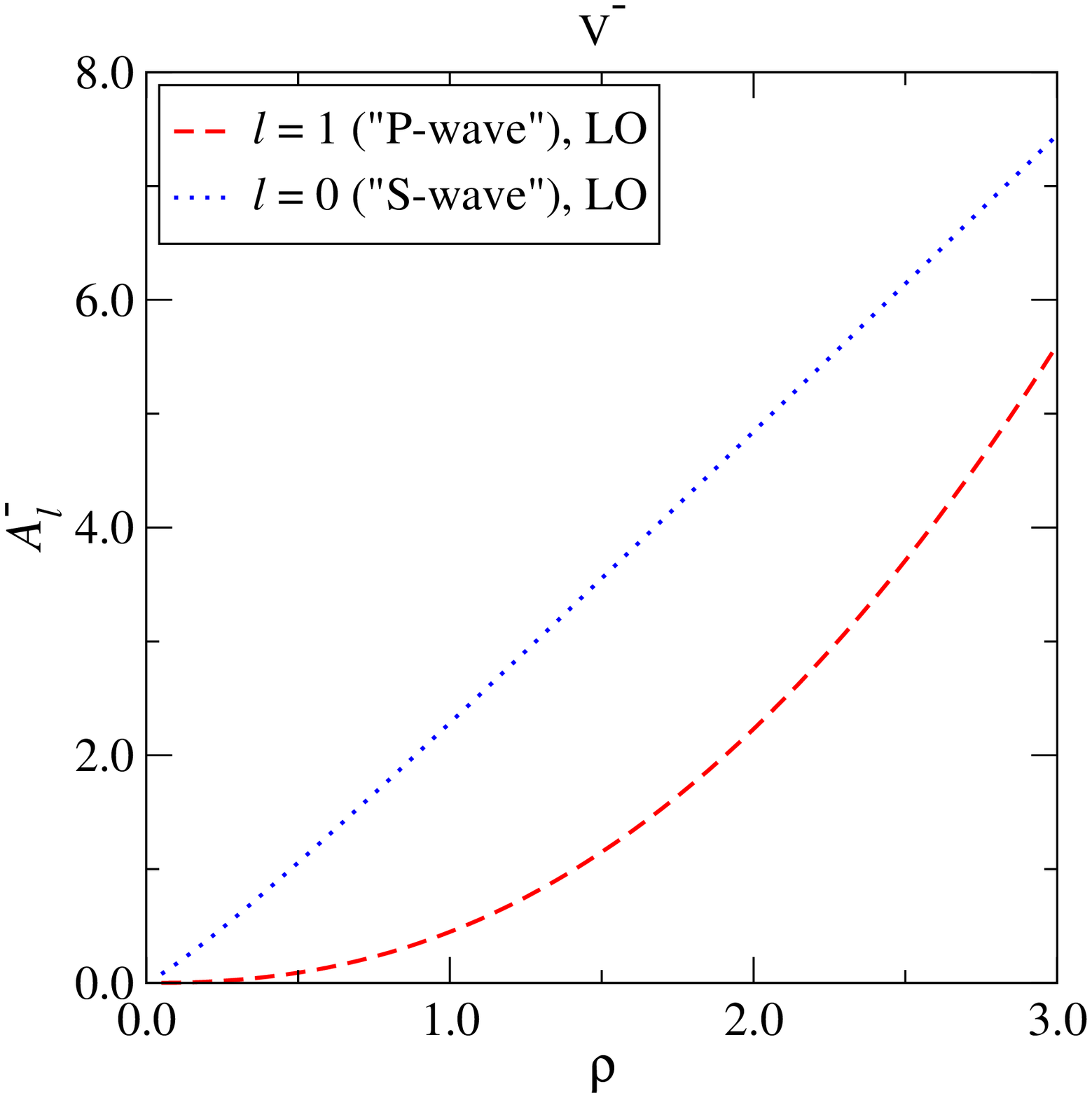}
}

\caption[a]{\small
Left: 
The lowest ``$S$-wave'' and ``$P$-wave'' 
eigenvalues obtained with 
the LO potential $V^{-}_\rmii{LO}$  (cf.\ \eq\nr{V_nlo_minus}). 
Right:
The ``amplitudes'' corresponding to the lowest eigenmodes 
(cf.\ \eq\nr{A_static}). 
}

\la{fig:E0hat_minus}
\end{figure}

We express the screening masses as ``energies'', 
\be
 E_\rmi{full} = M_\rmi{cm} + \frac{\gE^2\CF}{2\pi} \hat{E}
 \;, \la{Efull} 
\ee
and define the dimensionless quantities
\be
 \bar{y} \equiv \mE y \;, \quad
 \rho \equiv \frac{\gE^2\CF M_\rmi{r}}{\pi \mE^2}
 \;, \la{rho_def}
\ee
where $M_\rmi{r}$ is the reduced mass
(cf.\ \eqs\nr{M_kn1}, \nr{M_kn0}).
The radial homogeneous part of the
Schr\"odinger equation to be solved
(cf.\ \eqs\nr{g_eq}, \nr{f_eq}, \nr{g_eq_minus}, \nr{f_eq_minus}) reads
\be
 \biggl\{ - \frac{{\rm d}^2}{{\rm d}\bar{y}^2} - 
 \frac{1}{\bar{y}} \frac{{\rm d}}{{\rm d}\bar{y}} 
 + \frac{l^2}{\bar{y}^2} 
 + \rho \biggl( 
   \frac{2\pi V^\pm }{\gE^2 \CF} - \hat{E}^{(l)}
 \biggr) \biggr\} R_l = 0 
 \;, 
\ee
where $l=0,1,2,...$ denotes the angular quantum number. Assuming $R_l$
finite at $\bar{y} = 0$ and integrable at $\bar{y}\to\infty$, the 
eigenvalues $\hat{E}^{(l)}$ are easily determined numerically. Results
for the ground state ($\hat{E}^{(l)}_0$) 
for $V^{+}$ are shown in \fig\ref{fig:E0hat}(left) 
and for $V^{-}_{ }$ in \fig\ref{fig:E0hat_minus}(left).

Apart from the energies, the magnitudes of the correlators are
also of interest. These can be obtained by solving \eqs\nr{g_eq}, 
\nr{f_eq} in a spectral representation. Assuming a discrete spectrum
and letting $\psi^{ }_i$ be wave functions normalized as 
$
 \int \! {\rm d}^2\vec{y} \, \psi_i^*(\vec{y}) \psi^{ }_j(\vec{y}) =  
 \delta^{ }_{ij}
$, 
the solution of \eq\nr{g_eq} and subsequently \nr{g_limit} reads
\be
 g^{+}(\omega,\vec{y}) = \sum_{i=0}^{\infty} 
 \frac{\psi^{ }_i(\vec{y})\psi_i^*(\vec{0})}{E_i - \omega - i 0^+}
 \;, \quad
 \rho^{(k_n)}_{00} (\omega) = - 2 \pi \Nc T
 \sum_{i=0}^{\infty} \delta(E_i - \omega) |\psi^{ }_i(\vec{0})|^2
 \;, 
\ee
where
the sum $\sum_{0 < p_n < k_n}$ has been suppressed for notational
simplicity. 
Inserting this into \eq\nr{reconstruct}, we obtain 
the long-distance asymptotics
\be
 - \frac{G^{(k_n)}_{00}(z)}{T^3} \approx 
 \frac{\Nc \mE^2 \, \mathcal{A}^{+}_0 }{\pi T^2}
 \, e^{- |z| E_0^{(l=0)}} 
 \;, \quad
 \mathcal{A}^{+}_0 \equiv \frac{|R^{ }_0(0)|^2}
 {\int_0^\infty \! {\rm d}\bar{y} \, \bar{y} \, |R^{ }_0(\bar{y})|^2}
 \;. \la{A_0}
\ee
For the $P$-wave case, \eqs\nr{f_eq}, \nr{f_limit} lead similarly to 
\ba
 \vec{f}^{+}(\omega,\vec{y}) & = &  \sum_{i=0}^{\infty} 
 \frac{\psi^{ }_i(\vec{y})\nabla \psi_i^*(\vec{0})}{E_i - \omega - i 0^+}
 \;, \\
 \rho^{(k_n)}_{T} (\omega) & = & - \pi \Nc T 
 \biggl[  \frac{1}{p_n^2} + \frac{1}{(k_n - p_n)^2} \biggr]
 \sum_{i=0}^{\infty} \delta(E_i - \omega) | \nabla \psi^{ }_i(\vec{0}) |^2
 \;, 
\ea
and the configuration space correlator reads
(for $|z| \gg 1/[E_1^{(l=1)} - E_0^{(l=1)}]$)
\be
 - \frac{G^{(k_n)}_{T}(z)}{T^3} \approx
  \frac{\Nc \mE^4  \, \mathcal{A}^{+}_1}{\pi T^2 } 
 \biggl[  \frac{1}{p_n^2} + \frac{1}{(k_n - p_n)^2} \biggr]
 \, e^{- |z| E_0^{(l=1)}} 
 \;, \quad
 \mathcal{A}^{+}_1 \equiv \frac{|R'_1(0)|^2}
 {\int_0^\infty \! {\rm d}\bar{y} \, \bar{y} \, |R^{ }_1(\bar{y})|^2}
 \;. \la{A_1}
\ee
The ``amplitudes'' $\mathcal{A}^{+}_0, \mathcal{A}^{+}_1$ are illustrated
for the various potentials in \fig\ref{fig:E0hat}(right).

In the static sector, we similarly get from 
\eqs\nr{g_limit_minus} and \nr{f_limit_minus}
\ba
  - \frac{G^{(0)}_{T}(z)}{T^3} \approx
 \frac{4 \Nc \mE^2 \, \mathcal{A}^{-}_0 }{\pi T^2}
 \, e^{- |z| E_0^{(l=0)}} 
 \;, \quad
 - \frac{G^{(0)}_{00}(z)}{T^3} \approx
  \frac{4 \Nc \mE^4  \, \mathcal{A}^{-}_1}{\pi T^2 p_n^2 } 
 \, e^{- |z| E_0^{(l=1)}} 
 \;, \la{A_static}
\ea
where the eigenvalues and eigenfunctions are solved with $\hat{H}^-$, 
and the suppressed sum now reads
$ \sum_{p_n > 0} $.
The ``amplitudes'' $\mathcal{A}^{-}_0, \mathcal{A}^{-}_1$ are illustrated
in \fig\ref{fig:E0hat_minus}(right).

There is one more comment to make 
about the energies in \eq\nr{Efull}. For the non-static case, 
\eq\nr{M_kn1} implies
\be 
 E_0^{(l)} = M_\rmi{cm} + \frac{\gE^2 \CF}{2\pi} \hat{E}^{(l)}_0 
 = k_n + \frac{\gE^2 \CF}{2\pi} \biggl[
 \frac{\pi T}{4 M_\rmii{r}} + \hat{E}_0^{(l)} \biggr]
 \;.  \la{E_full_kn1}
\ee
Here we have re-expressed the parameter $m_\infty^2$ 
of \eq\nr{m_infty} in terms of the gauge coupling $\gE^2$. It should be 
noted however that $m_\infty^2$ is only known at 1-loop level whereas
the parameters $\mE^2$ and $\gE^2$ are known at 2-loop level. 
Within the same approximation, the ground state 
energies of the static sector are of the form
\be 
 E_0^{(l)} 
 = k_1 + \frac{\gE^2 \CF}{2\pi} \biggl[
 \frac{1}{2} + \hat{E}_0^{(l)} \biggr]
 \;.  \la{E_full_kn0}
\ee

%
\section{Lattice simulations}
\la{se:lattice}

%
\subsection{Basic setup}

For a non-perturbative crosscheck 
we make use of lattice simulations 
in two-flavour QCD, with physical parameters corresponding
to $\Lambdamsbar = 310(20)$~MeV 
and $m_\pi \approx 270$~MeV~\cite{T0}. Lattices of spatial size
$N_s^3 = 64^3$ and lattice spacing $a = 0.0486(4)(5)$~fm are considered. 
The thermal ensembles have 
temporal extents $N_\tau = 16, 12$, corresponding to the temperatures
$T = 254(4)$~MeV and $T=338(5)$~MeV, respectively. (In terms 
of the pseudocritical temperature of the QCD crossover these 
amount to $T/\Tc\simeq 1.2$ and $T/\Tc \simeq 1.6$ 
at $\Nf = 2$~\cite{Brandt:2013mba}.) 
Further details concerning the lattice setup and measurements
are given in appendix~B.

\begin{table}
\centerline{%
\begin{tabular}{|cllll|}
\hline
 $T/\Lambdamsbar$ &
 $\gE^2/T$ & 
 $\mE^2/T^2$ & 
 $\rho ^{(n = 0,1)}$ & 
 $\rho ^{(n = 2)}$ 
 \\
\hline
 0.82(5) & 3.2(2) & 3.5(4) & 0.61(4) & 0.92(9) 
  \\  
 1.09(7) & 2.8(2) & 3.1(3) & 0.61(2) & 0.91(4) 
  \\  
\hline
\end{tabular}}
\caption[a]{\small
 The effective gauge coupling, mass parameter, and $\rho$-parameter
 (cf.\ \eq\nr{rho_def})
 for the different sectors,  according to the 2-loop 
 computations in refs.~\cite{adjoint,gE2}.  
 The two temperatures correspond to 
 $T = 254(4)$~MeV  and
 $T = 338(5)$~MeV, respectively, for $\Nf = 2$ QCD. 
 The errors are based on variations of the renormalization scale.
  }
\la{table:eqcd}
\end{table}

\begin{table}[t]

\centerline{%
\begin{tabular}{|c|lll|lll|} 
\hline
 & \multicolumn{3}{c|}{$T/\Lambdamsbar = 0.82(5)$} 
 & \multicolumn{3}{c|}{$T/\Lambdamsbar = 1.09(7)$} \\
\hline
 & $n=0$ & $n=1$ & $n=2$  
 & $n=0$ & $n=1$ & $n=2$
  \\
\hline
  degeneracy & 
  2 & 1 & 2 & 2 & 1 & 2
  \\
 $E^{ }_{00} / T$ & 
 7.6(1)$^{\ast}$   & 8.0(2) & 14.0(1) 
 & 
 7.5(1)$^{\ast}$   & 7.8(1) & 13.8(1) 
 \\
 $E^{ }_{T} / T$ & 
 6.8(1)$^{\ast}$   & 9.2(2) & 15.0(2) 
 & 
 6.7(1)$^{\ast}$   & 8.9(2) & 14.7(1) 
  \\[1mm]
 $A^{ }_{00}/T^3$ & 
 0.7(2)$^{\ast}$   & 3.8(5) & 9.6(12) 
 & 
 0.5(1)$^{\ast}$   & 3.3(3) & 8.5(8)  
  \\
 $A^{ }_{T}/T^3$ & 
 17.8(22)$^{\ast}$ & 1.2(4) & 2.3(6) 
 & 
 15.7(14)$^{\ast}$ & 1.0(2) & 1.8(3) 
\\[2mm]
\hline 
\end{tabular}}

\caption[a]{\small
 Weak-coupling (for $n=0$, marked with an asterisk) and
 EQCD (for $n > 0$) predictions for screening masses and 
 ``amplitudes'', with the latter defined as $G \equiv - A\, e^{- E |z|}$
 at large $|z|$.
 For the amplitudes
 all states that are degenerate at the current level of precision have 
 been summed together.
 The errors, based on those in table~\ref{table:eqcd}, 
 should be considered as underestimates. 
  }
\la{table:pert}
\end{table}

\begin{table}[t]

\centerline{%
\begin{tabular}{|c|lll|lll|} 
\hline
 & \multicolumn{3}{c|}{$T/\mbox{MeV} = 254(4)$} 
 & \multicolumn{3}{c|}{$T/\mbox{MeV} = 338(5)$} \\
\hline
 & $n=0$ & $n=1$ & $n=2$  
 & $n=0$ & $n=1$ & $n=2$
 \\
 \hline 
  $E^{ }_{00} / T$ & 
 7.87(10)   & 7.45(6) & 13.6(4)$^{\ast}$ 
 & 
 7.69(23)   & 7.252(11) & 12.68(12)$^{\ast}$ 
 \\
 $E^{ }_{33} / T$ & 
 --   & 7.38(5) & 12.77(17)$^{\ast}$ 
 & 
 --   & 7.16(3) & 12.71(24)$^{\ast}$ 
 \\
 $E^{ }_{T} / T$ & 
 5.76(4)   & 9.35(20) & -- 
 & 
 6.097(12)   & 9.48(13) & -- 
  \\[1mm]
 $A^{ }_{00}/T^3$ & 
 7.9(7)   & 6.0(4) & 23.5(77)$^{\ast}$ 
 & 
 4.1(16)   & 4.78(7) & 15.8(16)$^{\ast}$  
  \\
 $A^{ }_{33}/T^3$ & 
 --   & 4.00(20) & 16.7(16)$^{\ast}$ 
 & 
 --   & 3.15(18) & 15.4(31)$^{\ast}$  
  \\
 $A^{ }_{T}/T^3$ & 
 10.3(4) & 9.2(16) & -- 
 & 
 10.63(13) & 10.0(12) & -- 
\\[2mm]
\hline 
\end{tabular}}

\caption[a]{\small
 Lattice results for the screening masses and 
 ``amplitudes''.  The errors are 
 statistical, with no estimate of systematics related 
 to cutoff effects. 
 For $n\neq 0$ 
 the screening masses $E^{ }_{00}$ and $E^{ }_{33}$
 should agree because of a Ward identity. 
 For $n=0$ the correlator $G^{ }_{33}$ is conserved
 and no screening mass can be extracted.
 For $n=2$ the data (marked with an asterisk)
 is noisy and the distances probed are close
 to the scale of the lattice spacing, so that systematic 
 uncertainties could be large. 
 We get no signal for the transverse correlator at $n=2$. 
  }
\la{table:lattice}
\end{table}

For a comparison with the results of the previous sections, the 
parameters of the effective theory need to be estimated. The 2-loop 
values for these, as well as for the parameter $\rho$ defined in 
\eq\nr{rho_def}, are given in table~\ref{table:eqcd}. The subsequent
predictions for the four-dimensional physical observables are 
shown in table~\ref{table:pert}. The energies come from 
\eqs\nr{E_full_kn1} and \nr{E_full_kn0}; the amplitudes 
from \eqs\nr{A_0}, \nr{A_1} and \nr{A_static}.

To the order we are working at, the ground state is degenerate in
several channels (cf.\ table~\ref{table:pert}). 
We expect this degeneracy to be lifted
at higher orders, in particular by the effect of the term $\delta S^{ }_0$
in the action (cf.\ \eq\nr{4quark}). 
In the practical lattice analysis we see no 
indications of closely lying states. Therefore, in the following, 
only single states are discussed on both sides.

%
\subsection{Fitting strategy}
\la{se:fitting}

In order to extract the screening masses and amplitudes from the 
non-perturbative lattice correlators, a fitting ansatz needs to be chosen.  
The discussion below refers to the form
\ba
 G_\rmi{cosh}(z)& \equiv & 
 A \, \frac{\cosh[M (z-L_z/2)]}{\sinh[M  L_z/2]}
 \;, \quad L_z \equiv 64 a \;, \la{Gcosh}
\ea
but we have also considered purely exponential fits of the form
$
 G_\rmi{exp}(z)  \equiv  \sum_{n=1}^2 
 A_n \, e^{- M_n z}
$.
The right edge of the fitting range is set to $L_z/2$, and the fits are 
repeated for all possible positions of the left edge. 
The results are extracted from uncorrelated fits with 
errors originating from a jackknife procedure. 
To decide which of the resulting parameters is the best we impose
a stability criterion with respect to the position of the left edge. 
To this end we compute the adjacent and next-to-adjacent
edge-position parameter values and demand that 
the difference of their average 
and the current parameters be smaller than some tolerance. 
In the next step we reduce this tolerance to the point 
where only a single parameter set fulfills the stability criterion,
and quote this number as our final result.

The qualities of such fits can be illustrated by defining ``effective
masses'' and ``effective amplitudes''. 
Effective masses are defined by the implicit equation
\be
 \frac{G(z-a/2)}{G(z+a/2)} =
 \frac{\cosh\big[ M_\rmi{eff}(z)( z-a/2 - {L_z}/{2}) \big]}
      {\cosh\big[ M_\rmi{eff}(z)( z+a/2 - {L_z}/{2}) \big]}
 \;.
 \la{effM}
\ee
In order to define effective amplitudes we divide the data by 
a function with the fitted mass value ($M_\rmi{fit}$) inserted
into \eq\nr{Gcosh}:
\be
 A_\rmi{eff}(z)=G(z)\,\frac{\sinh[M_\rmi{fit}L_z/2]}
 {\cosh[M_\rmi{fit}(z-L_z/2)]}
 \;. \la{effA}
\ee 
The results are shown for the lower temperature
$T=254(4)$~MeV in \fig\ref{fig:bare}. Like before, 
we refer to the screening masses in the following 
as ``energies'' ($M_\rmi{fit} \to E$).

\begin{figure}[t]


\centerline{%
 \epsfysize=7.8cm\epsfbox{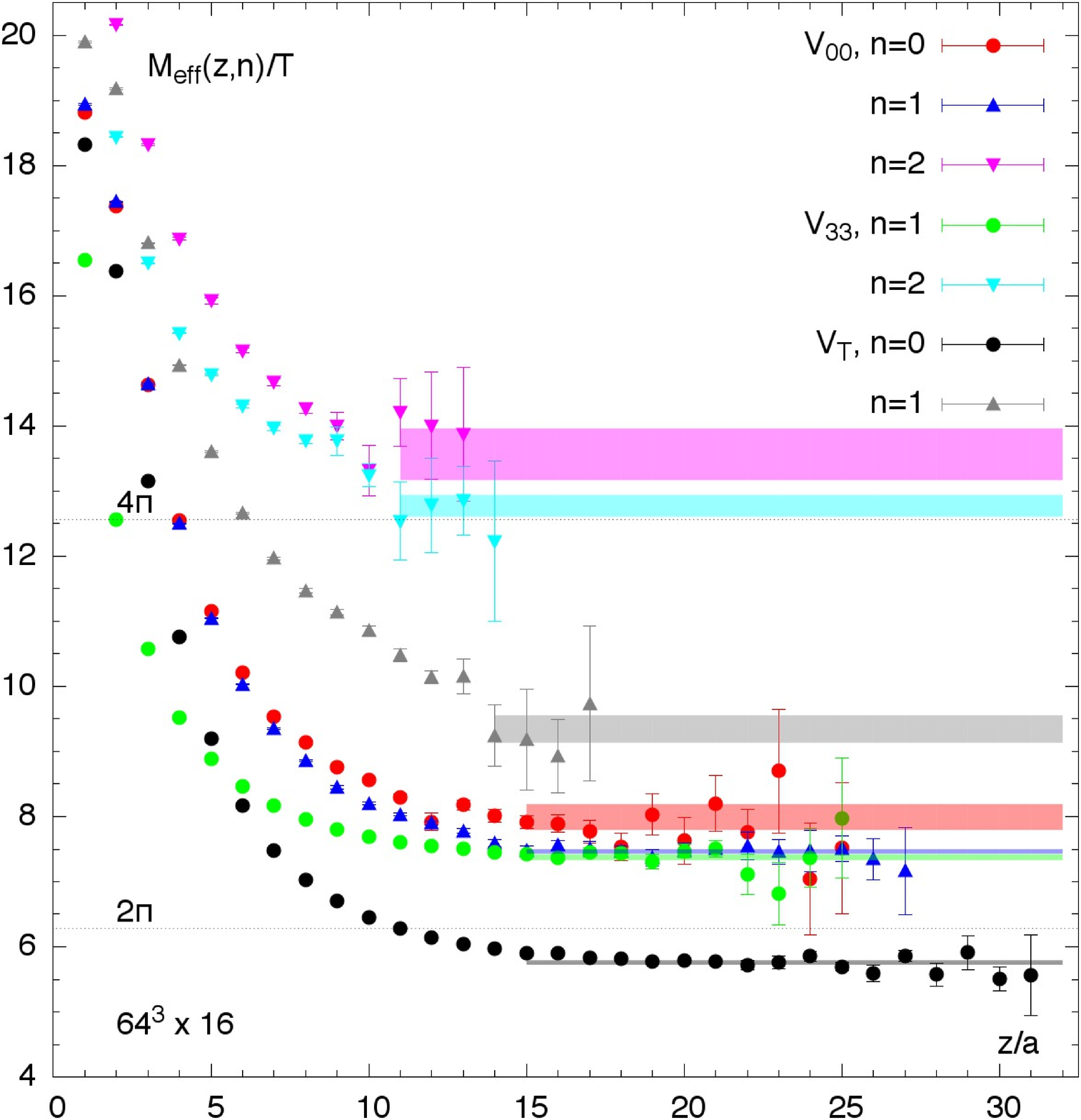}%
~~\epsfysize=7.8cm\epsfbox{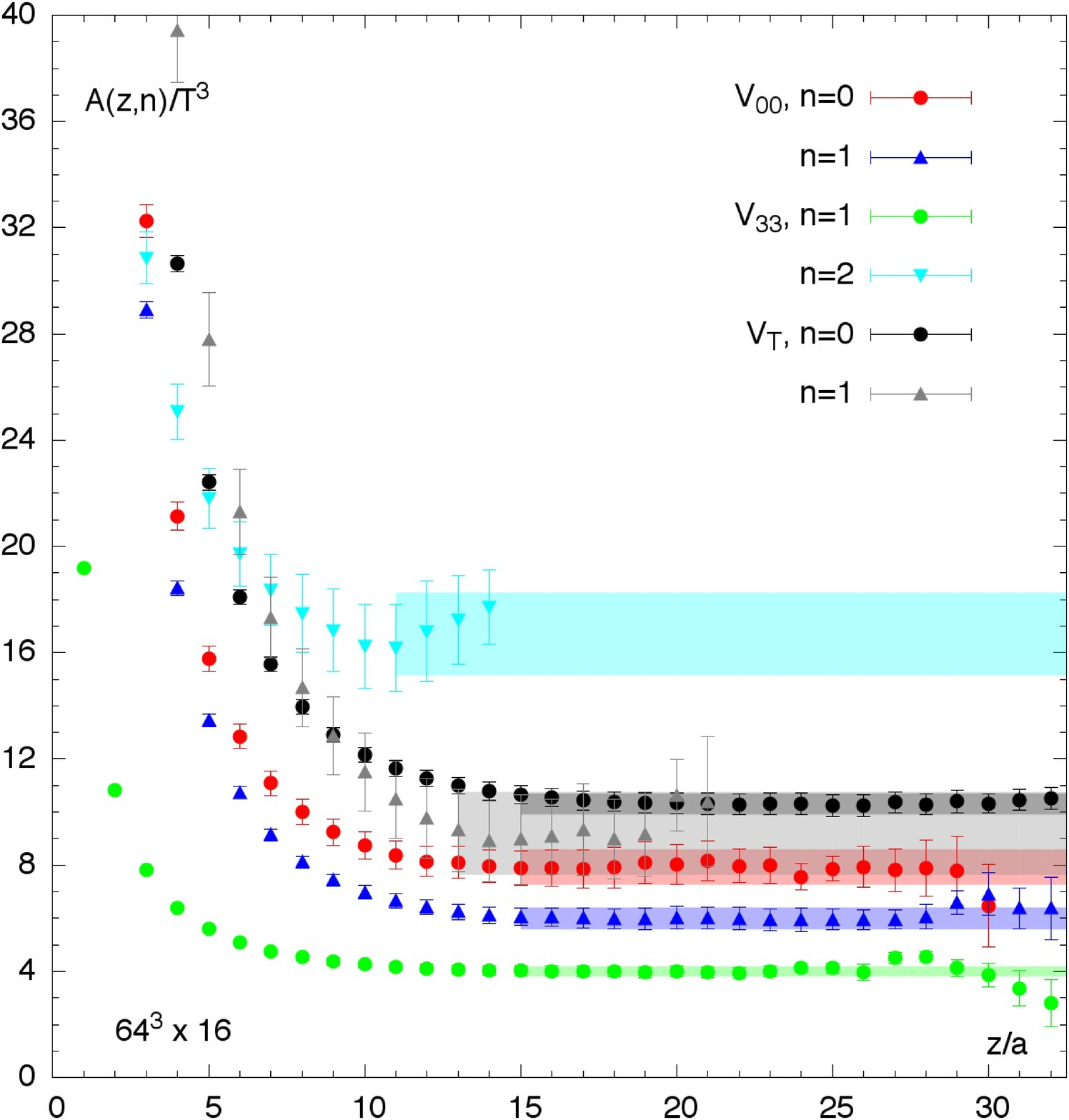}
}

\caption[a]{\small
Examples of effective masses (left) and effective amplitudes (right) 
for the lattice $16\times 64^3$ (the corresponding definitions are
given in \eqs\nr{effM}, \nr{effA}).
The results are collected in table~\ref{table:lattice}.
}

\la{fig:bare}
\end{figure}

For $n\neq 0$ a non-trivial crosscheck on the overall procedure 
can be obtained through Ward identities. Given the complex nature
of the fitting procedure and the fact that the lattice correlator
measured is of a local--conserved type (cf.\ appendix~B),  
the Ward identities are not trivially fulfilled. They assert
that $E^{(n=1)}_{00} = E^{(n=1)}_{33}$ and that  
$
 k_n^2 A_{00}^{(n=1)} = 
 [ E^{(n=1)}_{33} ]^2 A^{(n=1)}_{33} 
$. 
Using the average of 
$E^{(n=1)}_{00}$ and $E^{(n=1)}_{33}$ on the right-hand side we find 
for the ratio of the two sides
$1.08(10)$ at $T=254(4)$~MeV and $1.15(7)$ at $T=338(5)$~MeV, 
which indeed are consistent with unity within $\sim 2\sigma$ errors. 
The same consistency check is 
passed by the lattice data in the $n=2$ sector.

%
\subsection{Results and comparisons with perturbative predictions}

Our final results 
for the screening masses are shown in \fig\ref{fig:masses}, 
where they are also compared with the perturbative ones from 
table~\ref{table:pert}. The evolution of the perturbative
results in the non-static sector when going from 
LO to EQCD results is illustrated
in \fig\ref{fig:compare}. The final lattice results 
for the screening masses and amplitudes are collected in 
table~\ref{table:lattice}.

\begin{figure}[t]


\centerline{%
 \epsfysize=7.8cm\epsfbox{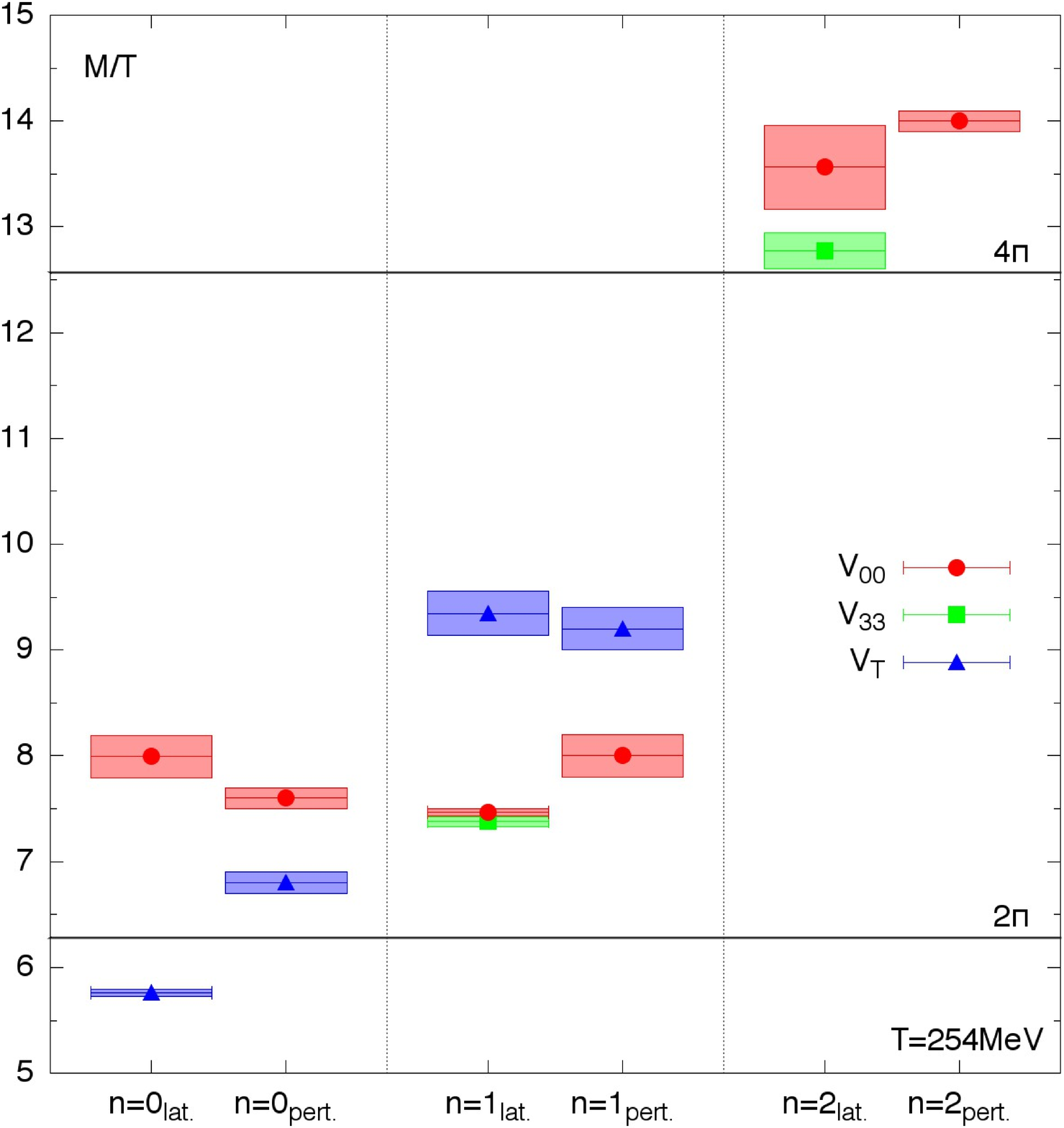}%
~~\epsfysize=7.8cm\epsfbox{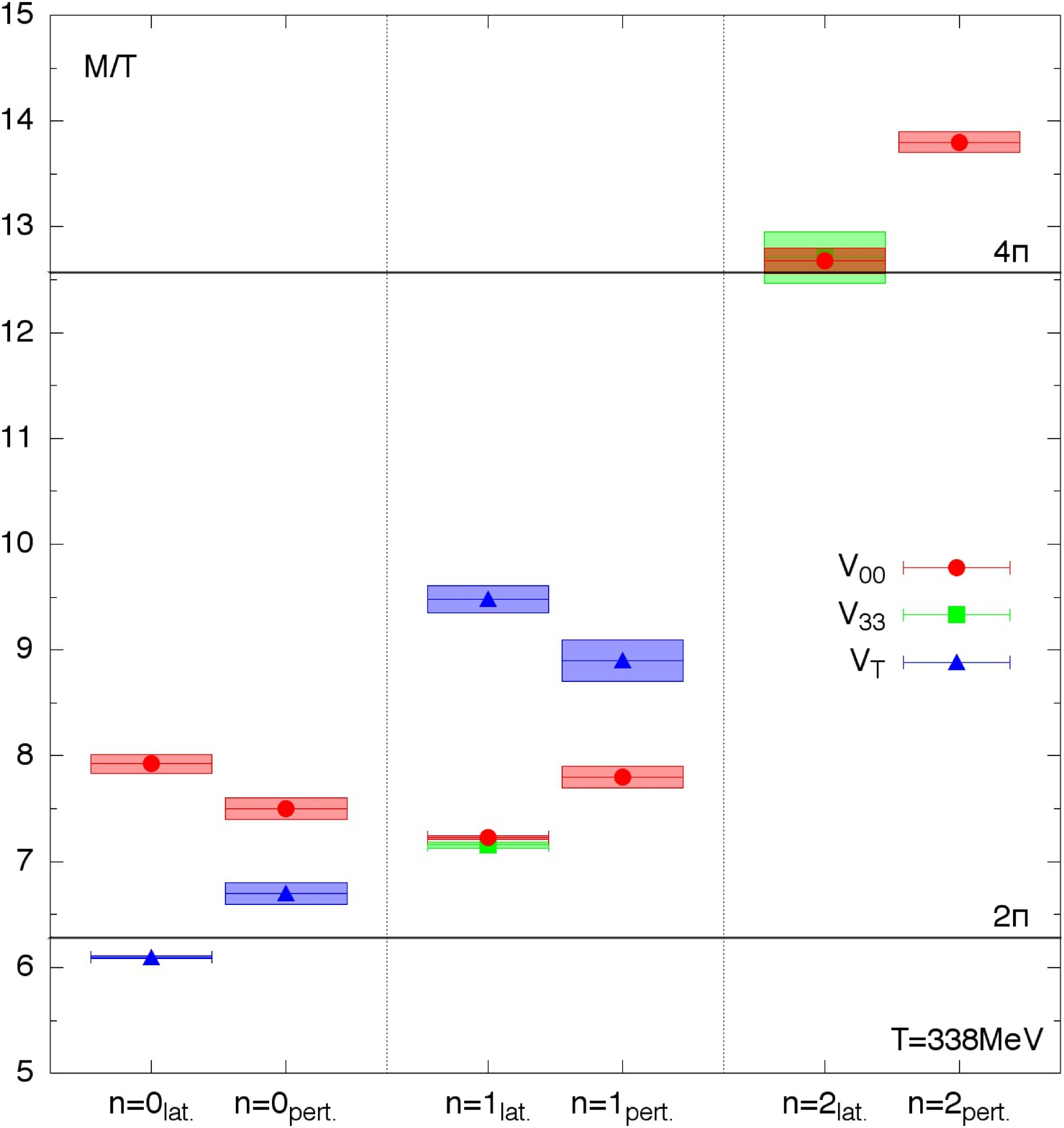}
}

\caption[a]{\small
The lattice screening masses at $T=254(4)$~MeV (left) and 
$T=338(5)$~MeV (right), compared with the corresponding weak-coupling
or effective-theory 
results from table~\ref{table:pert}. Note that 
discretization effects are expected to be 
larger at the higher temperature (i.e.\ $N_\tau = 12$), 
because then the distance scale 
$1/(4\pi T) = a N_\tau / (4\pi)$ is close
to the lattice spacing. 
}

\la{fig:masses}
\end{figure}

The following observations can be made:
\begin{itemize}

\item
On a rough level, the free-theory predictions 
$E^{(n=0)} = k_1$, $E^{(n > 0)} = k_n$ can be recognized 
in the full lattice data, 
with deviations that are $\lsim$ 50\%
(cf.\ \fig\ref{fig:masses}). 

\item
More quantitatively, 
for $n\neq 0$ the lattice and perturbative 
screening masses differ in general by less than 10\%
(cf.\ \fig\ref{fig:masses}). A fairly good 
agreement is also observed for the $P$-wave screening mass 
in the static sector ($E^{(n=0)}_{00})$.

\item
The $S$-wave screening masses in the static sector 
($E^{(n=0)}_T$) differ by about
15\% at the lower temperature, by 10\% at the higher one. 
Although not large per se, 
such discrepancies are beyond the 
estimated systematic errors of the effective description.
We recall, however, that in the static sector only the LO potential
is known, so that an additional approximation has been made: 
in the language of the introduction these predictions 
are of type (i) rather than (ii). 

\item
The splitting between the two $S$-wave masses 
($E^{(n=1)}_{00} - E^{(n=0)}_T$) 
is reproduced very well, particularly at the higher
temperature. It may be noted that at LO, the
splitting is entirely due to the difference in the sign of the force from 
the $A_0$ exchange, represented by the function $K_0^{ }(\mE y )$
in \eqs\nr{V_nlo} and \nr{V_nlo_minus}.

\item
The ``amplitudes'' do not compare well with each other: for the
$S$-wave cases ($A^{(n=0)}_{T}$, $A^{(n=1)}_{00}$)
the difference is $\sim 30\%$, but for the 
$P$-wave cases ($A^{(n=0)}_{00}$, $A^{(n=1)}_T$)
it is much larger (cf.\ tables~\ref{table:pert}, \ref{table:lattice}). 
For $n=2$ the difference is large 
even in the $S$-wave ($A^{(n=2)}_{00}$), and we get 
no signal in the $P$-wave ($A^{(n=2)}_T$).
The amplitudes may however
be expected to suffer from larger systematic uncertainties 
than the screening masses.
On the perturbative side, 
we have determined the correlators only to LO
as far as the amplitudes are concerned, not
to NLO like for the screening masses. 
Moreover, the LO result 
arises from a numerical solution of the wave function and is thereby 
sensitive to soft scales, yet for heavy states the effective 
theory description of the soft dynamics 
is likely to be less accurate than for the ground state.  
On the lattice side, the amplitudes
could be overestimated by misjudging where the plateau starts
(put another way, they could include contributions from almost 
degenerate excited states). In addition, discretization effects
have not been estimated, given that we only consider a single
lattice spacing. For all of these reasons we think that 
the $\sim 30\%$ discrepancy 
in the $S$-wave cases at $n=0,1$ is a reasonable 
reflection of the systematic uncertainties 
of a perturbative LO computation at the temperatures considered, 
whereas little can be deduced from the amplitudes related 
to the heavy states ($P$-wave or $n\ge 2$).   

\item
The $\sim 30\%$ resolution in the $S$-wave cases at $n=1$ 
($A_{00}^{(n=1)}$) could
only be reached thanks to the availability of the non-perturbative
EQCD potential. Had only the LO potential been available, the discrepancy
would have been $\sim 75\%$, cf.\ \fig\ref{fig:compare}(right).  

\item
Finally, we point out that in principle there is a ``two-meson'' 
threshold in each sector of fixed $n$. This physics has not been 
included in the current effective-theory analysis, 
however we believe that the states 
we have measured are light enough not to be affected.  

\end{itemize}

\begin{figure}[t]


\centerline{%
 \epsfysize=7.5cm\epsfbox{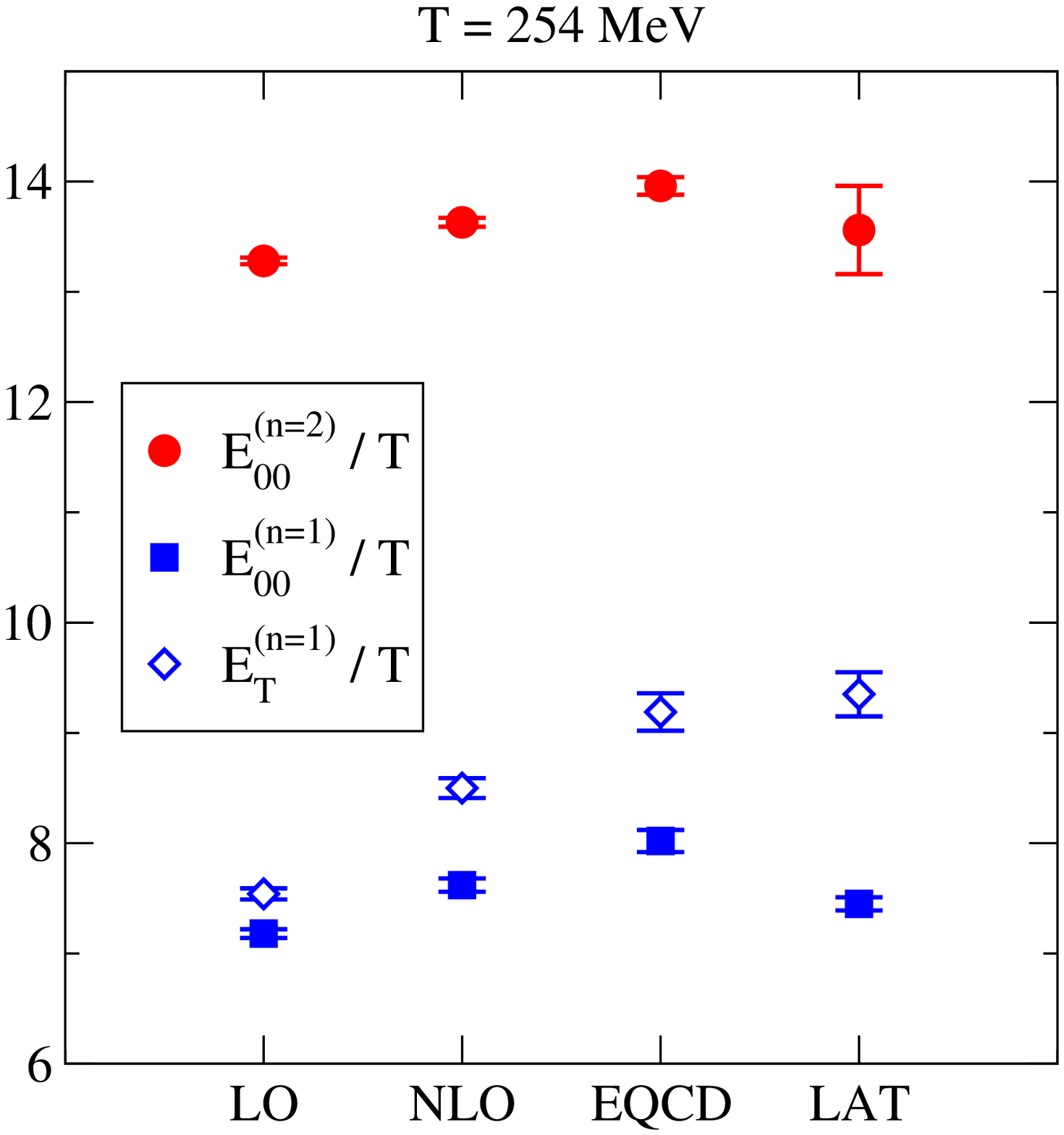}%
~~~\epsfysize=7.5cm\epsfbox{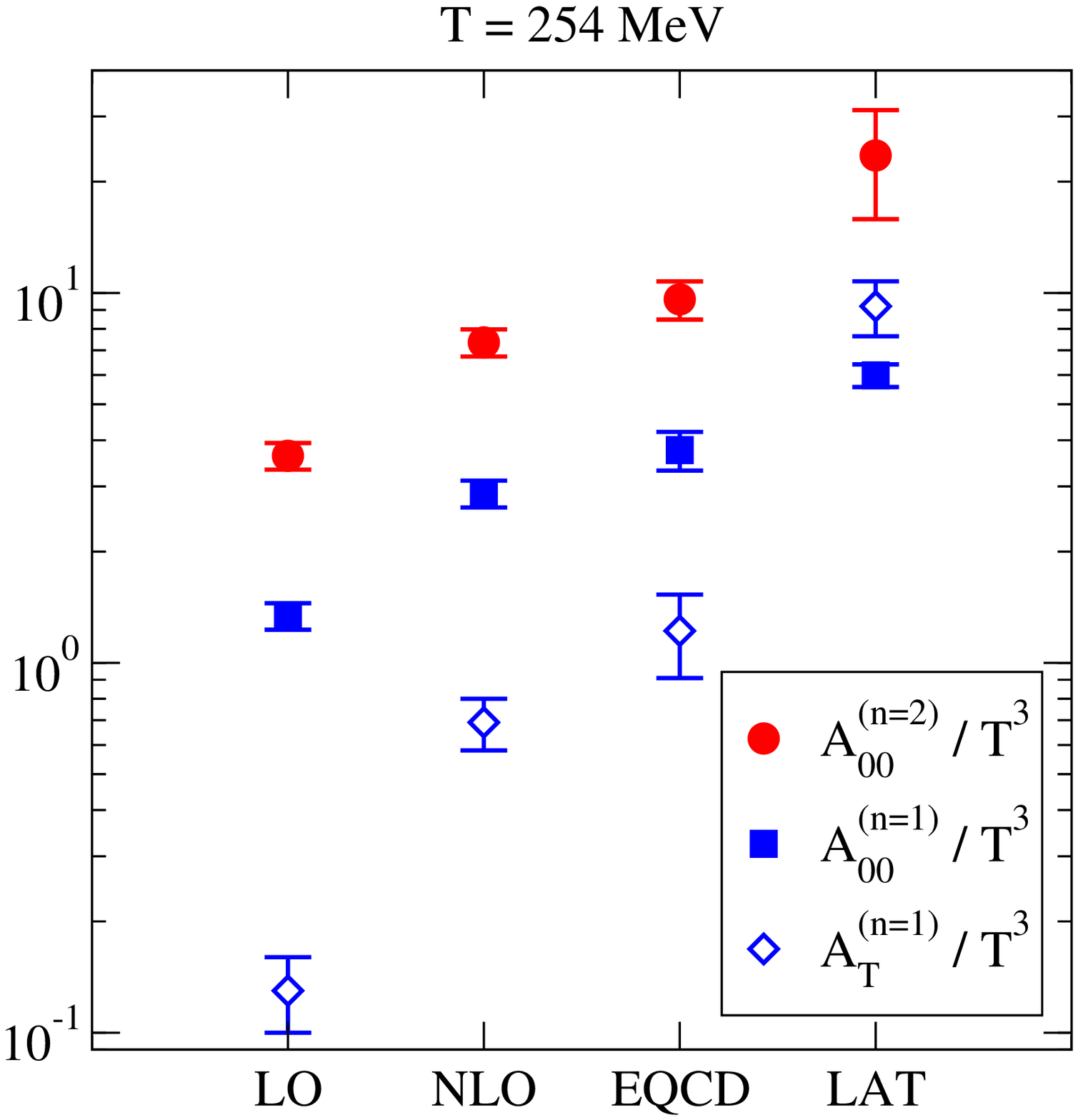}
}

\caption[a]{\small
Left: 
A comparison of the perturbative (LO, NLO, EQCD, referring to whether
the potential $V^+$ is from 
\eq\nr{V_nlo}, 
ref.~\cite{sch}, 
or ref.~\cite{marco}, respectively) 
and lattice (LAT)
results for the non-static screening masses at $T = 254$~MeV. 
Right: The same for the corresponding amplitudes. 
Going from LO to EQCD yields in most cases
an improvement, particularly for the amplitudes.
}

\la{fig:compare}
\end{figure}

\vspace*{-3mm}

%
\section{Conclusions}
\la{se:concl}

We have considered mesonic screening masses
related to the vector current both in a ``static''
(zero Matsubara frequency) and ``non-static'' (non-zero Matsubara 
frequency) sector. It turns out that even though in the weak-coupling 
limit both sectors probe physics at the momentum scale $gT$  and can 
be represented by ``non-relativistic'' low-energy effective theories  
in $2+1$ dimensions, 
the physical significances of the sectors are different. 
Indeed only the non-static sector has a clear relation 
to real-time physics: we have shown that the potential appearing 
in the effective description (cf.\ \eq\nr{V_nlo})
is identical to that previously considered
in the context of jet quenching and photon and dilepton production 
in thermal QCD (cf.\ e.g.\ refs.~\cite{sch}--\cite{agz_m}). 
On a general level, this observation is consistent
with the statement that only non-zero Matsubara modes play a role 
in analytic continuation from imaginary to real time~\cite{cuniberti}.

Apart from computing screening masses with an effective theory, 
we have also measured them with large-scale lattice Monte Carlo 
simulations in two-flavour QCD at temperatures of about 250~MeV
and 340~MeV. We find a remarkably good agreement
in the non-static sector, and also in the static sector 
for $P$-channel screening masses (cf.\ \fig\ref{fig:masses}). 
This adds confidence to the applicability of effective theory
methods for the study of phenomenologically interesting observables
in the temperature range relevant for heavy ion collision experiments. 
(For ``amplitudes'' the agreement
is poorer than for screening masses, but they also 
suffer from larger systematic errors both on the perturbative
and on the lattice side, as has been discussed 
around the end of \se\ref{se:lattice}.)

A general lesson, based on \fig\ref{fig:E0hat}, is that higher-order
corrections from the momentum scales $gT$ and $g^2 T/\pi$, even though 
formally suppressed by powers of $g$, are numerically of order 100\% compared
with results from the LO potential. 
In the case of the non-static sector we can at least partly 
account for these corrections, thanks to NLO computations and lattice
simulations carried out in the context of jet quenching~\cite{sch,marco}. 
Including these corrections in the solution 
of the Schr\"odinger equation indeed 
improves the overall agreement with lattice data
(cf.\ \fig\ref{fig:compare}). 

In the case of the static sector, in contrast, the 
potential is different and no NLO or non-perturbative results exist
for the moment. It seems conceivable, however, 
that computing such corrections
might permit to reduce discrepancies between weak-coupling
predictions and $S$-channel lattice data 
in the static sector. (The situation in the vector channel,  
illustrated in~\fig\ref{fig:masses}, is not too bad,
however the discrepancy is larger in the 
scalar and pseudoscalar channels, cf.\  
refs.~\cite{lat1,lat2} and references therein.) Part of the discrepancy
may be due to non-potential effects and spin-dependent terms, 
but it should be possible to incorporate these 
in the low-energy description as well.

In the present paper, we have demonstrated the existence of a relation
between screening masses and real-time rates through perturbative arguments, 
employing four-dimensional lattice simulations only as a crosscheck
for the accuracy of the perturbative description 
in the temperature range
considered. It would be very interesting if similar relations 
could be established on a non-perturbative level. As a modest
step in this direction, we may note that measuring the 
screening mass related to the operator 
$
 \int_0^{1/T} \! {\rm d}\tau \, e^{i k_n \tau} \, 
 \bar\psi (\tau, \tfr{\vec{y}}{2},z)
 \, \gamma_0 
 \,W^{ }_{\vec{y},z}\,
 \psi \bigl( \tau, - \tfr{\vec{y}}{2},z \bigr)  
$
directly in four dimensions, for $k_n$ large,  
and cancelling the free behaviour through an exponential
factor $\exp(k_n z)$ like in \eq\nr{defL}, 
would allow us to define fully non-perturbative variants of
the potential $V^+$ of \eq\nr{V_np}.

%
\section*{Acknowledgements}

We acknowledge the use of computing time on the JUGENE computer of 
the Gauss Centre for Supercomputing (FZ J\"ulich,
Germany); the $N_\tau=16$ lattice ensemble was generated within the
John von Neumann Institute for Computing (NIC) project HMZ21. The
$N_\tau=12$ ensemble was generated on the dedicated QCD platform
``Wilson'' at the Institute for Nuclear Physics, University of
Mainz, whereas the correlation functions were computed on the 
``Clover'' platform at the Helmholtz-Institute Mainz. 
The work of M.L was partly supported by the SNF 
under grant 200021-140234, and the work of H.M by the
  \emph{Center for Computational Sciences in Mainz} as part of the
  Rhineland-Palatinate Research Initiative and by the DFG under grant 
  ME 3622/2-1
 \emph{Static and dynamic properties of QCD at finite
    temperature}.

%
\appendix
\renewcommand{\thesection}{Appendix~\Alph{section}}
\renewcommand{\thesubsection}{\Alph{section}.\arabic{subsection}}
\renewcommand{\theequation}{\Alph{section}.\arabic{equation}}

%
\section{Higher modes ($|n| > 1$)}

In \se\ref{se:Seq}, results obtained from the effective theory description
were crosschecked against 
the free QCD results of \se\ref{se:full_lo} 
for $n = 0,1$. For completeness, we show here that the results match
also for a general $|n| > 1$.

Expanding \eqs\nr{G00_LO_z}, \nr{G11_LO_z} for a general $k_n$ 
at $z \gg 1 / |k_n|$, we obtain the asymptotics
\ba
 G^{(k_n)}_{00}(z) & = & -\Nc T^2 \,\frac{ e^{-|k_n| z}}{ z}
   \frac{2 \barkn^2 + 1}{6}
 + \rmO\Bigl ( \frac{ e^{- |k_n| z}}{z^2}  \Bigr)
 \;, \la{G00_LO_asy3} \\ 
 G^{(k_n)}_{T}(z)  
 & = & -\Nc T^2 \,  \frac{ e^{-|k_n| z}}{ z^2}
   \frac{4 \barkn^2 - 1}{3 |k_n| }
 + \rmO\Bigl ( \frac{ e^{-|k_n| z}}{z^3} \Bigr)
 \;.  \la{G11_LO_asy3}
\ea
The qualitative behaviours are the same as in 
\eqs\nr{G00_LO_asy}, \nr{G11_LO_asy2}, however there is a peculiar
dependence on $n$. This is related to a non-trivial 
``degeneracy'' of configurations  
leading to the same exponential fall-off at tree-level.

Consider a decomposition $k_n = p_n + (k_n - p_n)$, with $0 < p_n < k_n$.
By making use of the free value $M_\rmi{cm} = k_n$ (cf.\ \eq\nr{M_kn1}), 
\eqs\nr{rho00_nr} and \nr{rhoT_nr} read 
\ba
  \rho^{(k_n)}_{00,\rmii{LO}}(\omega) \!\! & = & \!\!
  - \Nc T \!\! \sum_{0 < p_n < k_n}
 \biggl( \frac{1}{p_n} + \frac{1}{k_n - p_n} \biggr)^{-1}
 \, \theta(\omega - k_n)
 \;, \\
  \rho^{(k_n)}_{T,\rmii{LO}}(\omega) \!\! & = & \!\!
  - \Nc T \!\! \sum_{0 < p_n < k_n}
 \biggl[ \frac{1}{p_n^2} + \frac{1}{(k_n - p_n)^2} \biggr]
 \biggl( \frac{1}{p_n} + \frac{1}{k_n - p_n} \biggr)^{-2}
 \!\!\! (\omega - k_n ) \, \theta(\omega - k_n)
 \;. \hspace*{5mm}
\ea
The sums can be carried out: 
\ba 
 \sum_{0 < p_n < k_n} \biggl( \frac{1}{p_n} + \frac{1}{k_n - p_n} \biggr)^{-1}
 & = & \pi T \, \frac{2\barkn^2 + 1}{ 6} \;, 
 \\ 
 \sum_{0 < p_n < k_n} 
 \biggl[ \frac{1}{p_n^2} + \frac{1}{(k_n - p_n)^2} \biggr]
 \biggl( \frac{1}{p_n} + \frac{1}{k_n - p_n} \biggr)^{-2}
 & = & \frac{4\barkn^2 - 1}{ 6 \barkn } \;. 
\ea
We observe the same prefactors as in the full QCD 
results of \eqs\nr{G00_LO_asy3}, \nr{G11_LO_asy3}. Carrying 
out the Laplace transform in \eq\nr{reconstruct} the overall
coefficients can be seen to agree as well.  

%
\section{Technical details related to lattice simulations}

Our simulations are based on 
the standard Wilson gauge action, with fermions implemented 
via the O($a$) improved Wilson discretization with a non-perturbatively 
determined clover coefficient $c_\rmi{sw}$~\cite{Jansen:1998mx}.  
The configurations were generated with the MP-HMC
algorithm~\cite{Hasenbusch:2001ne,Hasenbusch:2002ai} 
employing the implementation of ref.~\cite{Marinkovic:2010eg} 
based on 
L\"uscher's DD-HMC package~\cite{CLScode}. 

Spatial correlation functions were computed
on two ensembles, using the same discretization and
masses as in the sea sector. The first ensemble, with a spatial
size $N_s^3=64^3$ and a temporal extent of $N_\tau=16$, consisted 
of 313 independent configurations. It was first
presented in ref.~\cite{Tn0} and has subsequently been analyzed in
refs.~\cite{lat0,Brandt:2013fg}. The second ensemble is newly generated,
and has 262 configurations on an $N_\tau \times N_s^3=12\times 64^3$
lattice.   
Both ensembles were generated at fixed bare parameters, 
corresponding to a lattice spacing 
$a=0.0486(4)(5)$fm~\cite{T0} so that $a m_\pi N_s  = 4.2$. Inserting into
$T=1/(N_\tau a)$ the two ensembles correspond to $T=254(4)$~MeV at
$N_\tau=16$ and $T=338(5)$~MeV at $N_\tau=12$. 

As in ref.~\cite{lat0}, we implemented 
the vector correlation function as a mixed correlator 
between a local and a conserved current. The three correlators
considered are
\ba 
 G_T^{(k_n)\rmi{bare}}(z)
 & = & - a^3 
 \sum_{i,\tau,\vec{x}} 
 \, e^{i k_n \tau} \,
 \big\langle \, J^c_i(\tau,\vec{x},z)\, J^l_i({0}) \, 
 \big\rangle \;, \la{GTbare} \\
 G_{00}^{(k_n)\rmi{bare}}(z) 
 & = & - a^3 \sum_{\tau,\vec{x}}
 \, e^{i k_n \tau} \,
 \big\langle \, J^c_0(\tau,\vec{x},z)\, J^l_0({0}) \,
 \big\rangle \;, \\
 G_{33}^{(k_n)\rmi{bare}}(z) 
 & = & + a^3 \sum_{\tau,\vec{x}} 
 \, e^{i k_n \tau} \,
 \big\langle \, J^c_3(\tau,\vec{x},z)\, J^l_3({0}) \,
 \big\rangle
 \;, 
 \la{G33bare}
\ea
where minus signs have been inserted in order to obtain positive 
correlators. The local ($l$) and conserved ($c$) 
currents are defined as ($x \equiv (\tau,\vec{x},z)$) 
\ba
 J_\mu^l({x}) \! & \equiv & \!
 \frac{1}{\sqrt{2}} \,
 \bar q({x})\, \gamma_\mu {\sigma^{ }_3}\, q({x})
 \;, 
 \\
 J_\mu^c ( {x} ) \! & \equiv & \!
 \frac{1}{2\sqrt{2}} \,
 \Big[ 
  \bar q ( {x} + a\hat\mu )\, ( 1 + \gamma_\mu ) U_\mu^\dagger ( {x} ) 
 {\sigma^{ }_3}\, q ( x ) 
 - \bar q ( {x} )\, ( 1 - \gamma_\mu ) U_\mu ( {x} )
  {\sigma^{ }_3}\, q ( {x} + a\hat\mu ) 
 \Big]
 \;. \hspace*{6mm}   \la{eq:Jdef}
\ea
Here $q$ represents a mass-degenerate quark doublet, $\sigma^{ }_3$ 
a diagonal Pauli matrix acting on flavour indices, and $U_\mu$
a link matrix. The doublet can be
interpreted as the $u$ and $d$ quarks. 

In order to enhance the statistical precision of the measurements, 
we supplement the standard source at position $x_\rmi{src}=0$ with 
$N_\rmi{src}=64$ additional randomly chosen source positions in 
the lattice four-volume, thus obtaining $\lsim 1\%$
statistical errors for the $S$-wave masses. 

The local (non-conserved) vector current $J_\mu^l$ requires a finite
renormalization factor $Z_\rmii{V}$ (cf.\ e.g.\ ref.~\cite{uk}).  
Correspondingly the bare vector correlators were renormalized using
\be
 G_{\mu\nu}^{(k_n)}(z)= Z_\rmii{V}(g_0^2)\,  G_{\mu\nu}^{(k_n)\rmi{bare}}(z) 
 \;, \la{renorm}
\ee
with the non-perturbative value 
$Z_\rmii{V}(g_0^2) =0.768(5)$ at 
$6/g_0^2 = 5.50$~\cite{DellaMorte:2005rd}.  We have not included O($a$)
contributions from the improvement term proportional 
to the derivative of the antisymmetric tensor
operator~\cite{Luscher:1996sc,Sint:1997jx}, nor
a quark-mass dependent improvement  
of the form $1+b_\rmii{V}(g_0^2) a m_q$~\cite{Sint:1997jx}.
Both should be included to ensure a smooth 
scaling behaviour as the continuum limit is taken, however our
present study concerns a single (fine) lattice spacing.

\begin{figure}[t]


\centerline{%
 \epsfysize=7.8cm\epsfbox{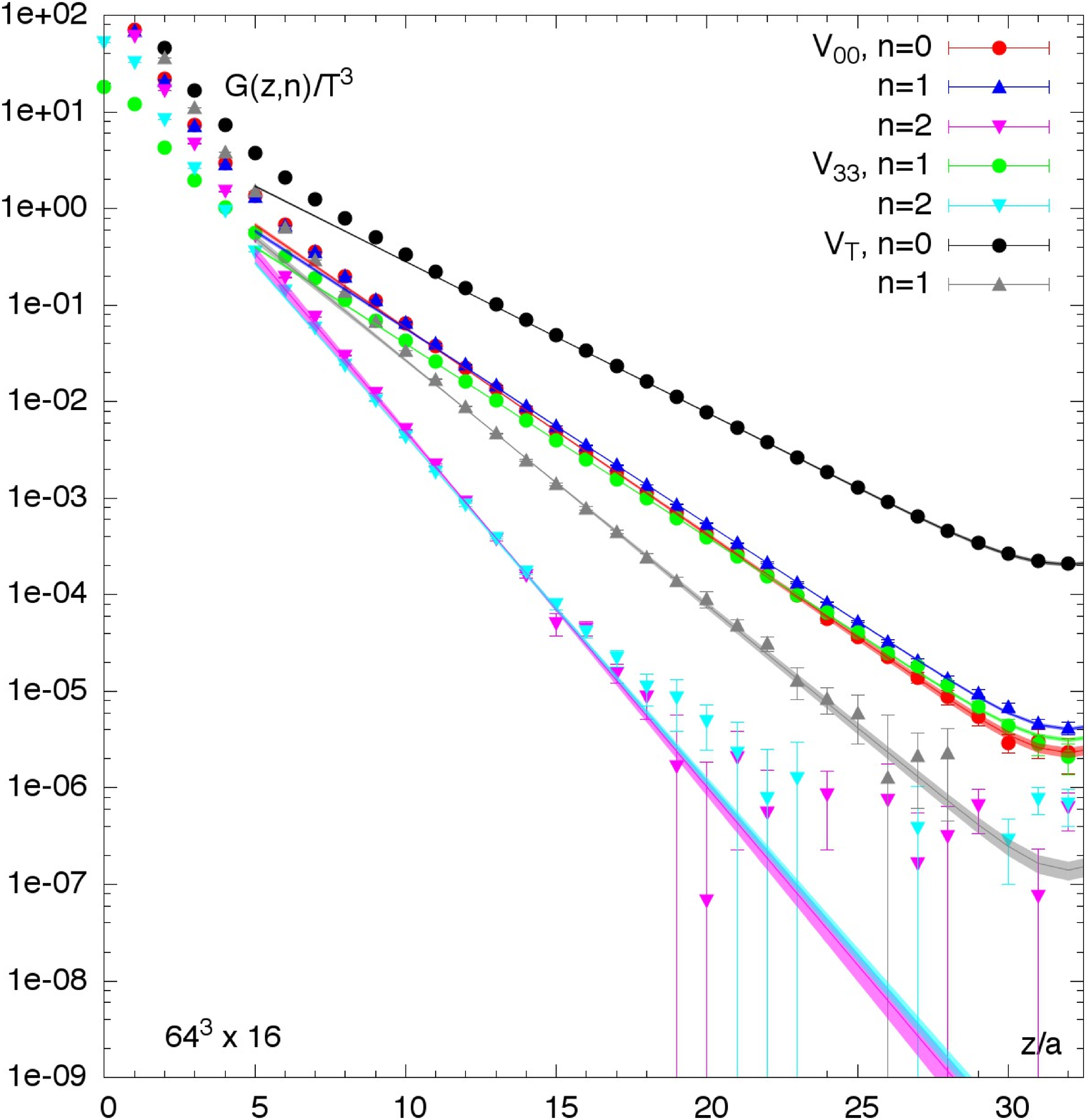}%
~~\epsfysize=7.8cm\epsfbox{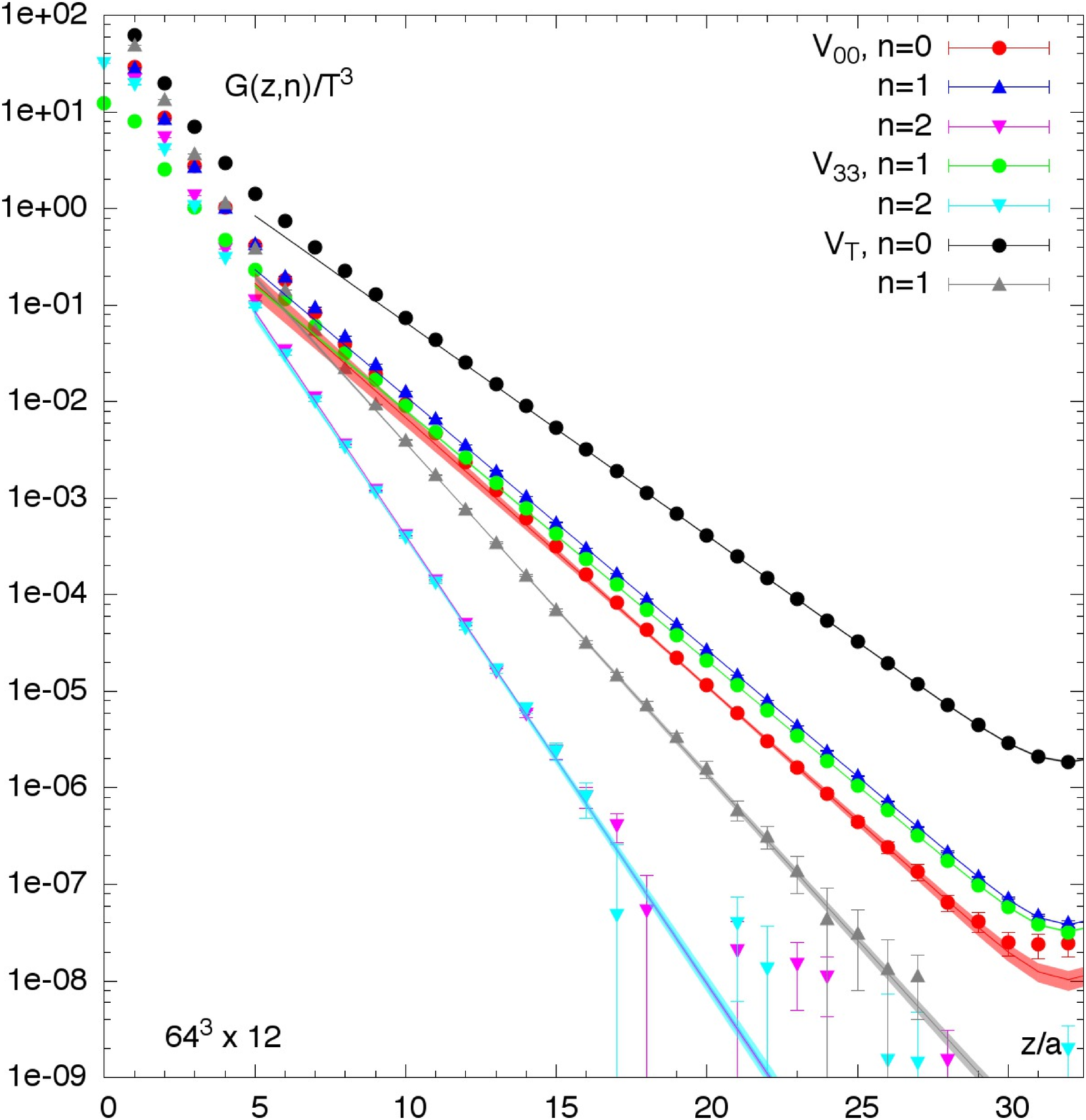}
}

\caption[a]{\small
 The correlators defined in \eqs\nr{GTbare}--\nr{G33bare}, 
 renormalized according to \eq\nr{renorm}, at
 $T=254(4)$~MeV (left panel) and $T=338(5)$~MeV (right panel). 
 The shaded bands represent the fitted results with 
 the corresponding ground state masses.
}

\la{fig:correlators}
\end{figure}

The renormalized lattice data as well as the fitted correlation functions 
for the ground states are shown as the coloured shaded bands 
in \fig\ref{fig:correlators}. 
The error estimates were obtained via a jackknife procedure. 


\end{document}